\documentclass[aps,prd,nofootinbib,twocolumn,showpacs,superscriptaddress,letterpaper]{revtex4-2}
\pdfoutput=1 

\usepackage[T1]{fontenc}
\RequirePackage[running,mathlines]{lineno} 
\usepackage{bm,amsmath,amsfonts,amssymb,wasysym}
\usepackage{graphicx}
\usepackage{breakurl}
\usepackage{array}
\usepackage{slashed}
\usepackage{xspace}
\usepackage{hhline}
\usepackage[italic]{hepnames}
\usepackage{color}
\usepackage{placeins}
\usepackage{adjustbox}
\usepackage{rotating} % Rotating table
\usepackage{tabularx} 
\usepackage[vlines]{tabularht}
\usepackage[separate-uncertainty=true]{siunitx}
\usepackage{physics}
\usepackage{gensymb}
\usepackage{hyphenat}
\usepackage{subfigure}
\usepackage{url}

\usepackage{booktabs}
\usepackage{multirow}
\usepackage{makecell}

\usepackage{lipsum}
\usepackage{mathtools}
\usepackage{cuted}
\usepackage{supertabular}
\usepackage{soul}

\usepackage[colorlinks,hyperindex,breaklinks]{hyperref}
\hypersetup{linkcolor=blue,citecolor=blue,filecolor=black,urlcolor=blue}

\usepackage[capitalise]{cleveref}

\usepackage{orcidlink}
\usepackage{adjustbox}
\usepackage{rotating}  
\usepackage{dcolumn}       % <--- added
\usepackage{tablefootnote}

\def\lumi{\ensuremath{189 \, \mathrm{fb}^{-1}}\xspace}
\def\NBB{\ensuremath{ N_{B {\kern 0.18em\overline{\kern -0.18em B}}} = \left(198 \pm 3 \right) \times 10^{6}}\xspace}

\def\bdslnu{\ensuremath{{\kern 0.18em\overline{\kern -0.18em B}}{}^0\to D^{*+}\ell^-\bar\nu_\ell}\xspace}

\def\bdsenu{\ensuremath{{\kern 0.18em\overline{\kern -0.18em B}}{}^0\to D^{*+}e^-\bar\nu_e}\xspace}
\def\bdsmunu{\ensuremath{{\kern 0.18em\overline{\kern -0.18em B}}{}^0\to D^{*+}\mu^-\bar\nu_\mu}\xspace}

\def\Bzbar{\ensuremath{{\kern 0.18em\overline{\kern -0.18em B}}{}^0}\xspace}

\def\DstDpi{\ensuremath{D^{*+}\to D^0 \pi^+}\xspace}
\def\DKpi{\ensuremath{D^0\to K^-\pi^+}\xspace}

\def\GF{\ensuremath{G_{F}}\xspace}

\def\b2{Belle~II\xspace}

\def\arraystretch{1.3}

\def\Bbar{\kern 0.18em\overline{\kern -0.18em B}\xspace}
\newcommand{\cosBY}{\ensuremath{\cos\theta_{BY}}\xspace}
\newcommand{\DeltaM}{\ensuremath{\Delta  M}\xspace}
\newcommand{\GeV}{GeV/\ensuremath{c^2 }\xspace}

\newcommand{\Vcb}{\ensuremath{|V_{cb}|}\xspace}
\newcommand{\VcbBGL}{\ensuremath{|V_{cb}|_\mathrm{BGL}=(40.57\pm 0.31 \pm 0.95\pm 0.58 )\times 10^{-3}}\xspace}
\newcommand{\VcbCLN}{\ensuremath{|V_{cb}|_\mathrm{CLN}=(40.13 \pm 0.27 \pm 0.93\pm 0.58 )\times 10^{-3}}\xspace}
\newcommand{\DAFB}{\ensuremath{\Delta{\cal A}_{\mathrm{FB}} = (-17 \pm 16 \pm 16)\times 10^{-3}}\xspace}
\newcommand{\DFL}{\ensuremath{\Delta F_L = 0.006 \pm 0.007 \pm 0.005}\xspace}
\newcommand{\Remu}{\ensuremath{0.998 \pm 0.009 \pm 0.020}\xspace}
\newcommand{\BR}{\ensuremath{{\cal B}(\bdslnu)=(4.922 \pm 0.023 \pm 0.220)\%}\xspace} 

\newcommand{\kinematic}{$w$, $\cos\theta_\ell$, $\cos\theta_V$, and $\chi$\xspace}

\begin{document}

\vspace*{-5\baselineskip}
\begin{flushright}
Belle~II Preprint 	2023-014		\\	
KEK Preprint 		2023-28 \\
\end{flushright}

\title{Determination of $|V_{cb}|$ using \bdslnu  decays with Belle~II}

%%%%%%%%%%%%%%%%%%%%%%%%%%%%%%%%%%%

% \author{F.~Abudin{\'e}n\,\orcidlink{0000-0002-6737-3528}} % 2250
  \author{I.~Adachi\,\orcidlink{0000-0003-2287-0173}} % 2590
% \author{K.~Adamczyk\,\orcidlink{0000-0001-6208-0876}} % 2239
  \author{L.~Aggarwal\,\orcidlink{0000-0002-0909-7537}} % 10083
% \author{P.~Ahlburg\,\orcidlink{0000-0002-9832-7604}} % 2367
  \author{H.~Ahmed\,\orcidlink{0000-0003-3976-7498}} % 11323
% \author{J.~K.~Ahn\,\orcidlink{0000-0002-5795-2243}} % 7423
  \author{H.~Aihara\,\orcidlink{0000-0002-1907-5964}} % 2223
  \author{N.~Akopov\,\orcidlink{0000-0002-4425-2096}} % 9443
  \author{A.~Aloisio\,\orcidlink{0000-0002-3883-6693}} % 2194
% \author{L.~Andricek\,\orcidlink{0000-0003-1755-4475}} % 2098
  \author{N.~Anh~Ky\,\orcidlink{0000-0003-0471-197X}} % 2218
  \author{D.~M.~Asner\,\orcidlink{0000-0002-1586-5790}} % 4684
  \author{H.~Atmacan\,\orcidlink{0000-0003-2435-501X}} % 2538
% \author{V.~Aulchenko\,\orcidlink{0000-0002-5394-4406}} % 8183
  \author{T.~Aushev\,\orcidlink{0000-0002-6347-7055}} % 3747
% ...
% 2387
% \author{V.~Zhulanov\,\orcidlink{0000-0002-0306-9199}} % 4983
% \author{R.~\v{Z}leb\v{c}\'{i}k\,\orcidlink{0000-0003-1644-8523}} % 13403
\collaboration{The Belle II Collaboration}

\noaffiliation
%%%%%%%%%%%%%%%%%%%%%%%%%%%%%%%%%%%

\begin{abstract}
We determine the Cabibbo-Kobayashi-Maskawa matrix-element magnitude \Vcb using \bdslnu decays reconstructed in \mbox{\lumi} of collision data collected by the Belle~II experiment, located at the SuperKEKB $e^+e^-$ collider. Partial decay rates are reported as functions of the recoil parameter $w$ and three decay angles separately for electron and muon final states. We obtain \Vcb using the Boyd-Grinstein-Lebed  and Caprini-Lellouch-Neubert parametrizations, and find \VcbBGL and \VcbCLN with the uncertainties denoting statistical components, systematic components, and components from the lattice QCD input, respectively. The branching fraction is measured to be \BR. The ratio of branching fractions for electron and muon final states is found to be \Remu. In addition, we determine the forward-backward angular asymmetry and the $D^{*+}$ longitudinal polarization fractions. All results are compatible with lepton-flavor universality in the Standard Model.
\end{abstract}

 
\maketitle

%%%%%%%%%%%%%%%%%%%%%%%%%%%%%%%%%%%
\section{Introduction}\label{sec:introduction}

In the Standard Model (SM) of particle physics, the Cabibbo-Kobayashi-Maskawa (CKM) matrix relates the weak interaction and mass eigenstates of quarks~\cite{PhysRevLett.10.531,10.1143/PTP.49.652}. It is a $3 \times 3$ unitary complex matrix and currently the only established source for charge-parity ({\it CP}) violating processes in the SM. The elements of the CKM matrix are free parameters of the SM and need to be determined experimentally.
In this paper, we report the determination of the magnitude of the CKM matrix element $V_{cb}$ with Belle~II data. The values of \Vcb should be consistent when determined using different physics processes involving $b \to c$ quark transitions. However, a longstanding discrepancy between inclusive and exclusive determinations is observed, cf. Ref.~\cite{Workman:2022ynf}. Reference~\cite{HFLAV:2022pwe} combines measurements of exclusive $B \to  D^{(*)} \ell \bar \nu_\ell$ and $B^0_s \to D_s^{(*)-} \, \ell^+ \nu_\ell$ decays, where $\ell = e, \mu $, and $B$ and $D$ indicate charged and neutral bottom and charmed mesons, respectively. This combination results in a value of
\begin{linenomath*}
\begin{equation}
    |V_{cb}|=(39.10 \pm 0.50)\times 10^{-3} \quad (\text{exclusive}).
\end{equation}
\end{linenomath*}
The most precise inclusive determination combines measured spectral moments of lepton energy and hadronic invariant-mass spectra with measurements of partial branching fractions and calculations of the total rate at next-to-next-to-next-to-leading order in the strong interaction~\cite{Bordone:2021oof},  
\begin{linenomath*}
\begin{equation}
    |V_{cb}|=(42.16 \pm 0.51)\times 10^{-3} \quad (\text{inclusive}).
\end{equation}
\end{linenomath*}
Alternative inclusive determinations obtain a similar value using lepton-mass squared moments~\cite{Fael:2018vsp, Bernlochner:2022ucr, Belle-II:2022fug, Belle:2021idw}. 

In this paper, we examine the semileptonic \bdslnu decay,\footnote{Charge-conjugate modes are implicitly included throughout this paper.} which is a usual benchmark channel to determine \Vcb due to the sizeable branching fraction, small backgrounds, and the availability of lattice QCD (LQCD) information at and beyond zero recoil~\cite{FermilabLattice:2014ysv,FermilabLattice:2021cdg, Aoki:2023qpa, Harrison:2023dzh}. LQCD inputs aid in disentangling the normalization of the hadronic-transition form factors, which describe the nonperturbative processes of the strong interaction in the $B \to D^{(*)}$ transition. 

Aside from the \Vcb discrepancies, several observations suggest a possible deviation from the SM's lepton-flavor universality (LFU) in semileptonic $B$ decays (see, e.g., ~\cite{Crivellin:2021sff,Bernlochner:2021vlv} for recent reviews). For example, the combined averages of ${\cal R}(D)$ and ${\cal R}(D^{*})$ are about three standard deviations larger than the SM expectation~\cite{HFLAV:2022pwe}, where
\begin{linenomath*}
\begin{equation}
 {\cal R}(D^{(*)})=\frac{{\cal B}(B\rightarrow D^{(*)}\tau \bar\nu_\tau)}{{\cal B}(B\rightarrow D^{(*)}\ell \bar\nu_\ell )}, \quad \text{with } \ell=e,\, \mu\, .
\end{equation}
\end{linenomath*}
This ratio does not depend on \Vcb, but the most precise predictions rely on measurements of the form factors. In addition, a reanalysis~\cite{Bobeth:2021lya} of Belle data~\cite{Belle:2018ezy} recently extracted the difference of the forward-backward angular asymmetry between electrons and muons in \bdslnu decays
and observed a significant tension of about four standard deviations with the SM expectation. Motivated by this, the Belle~II collaboration carried out tests of LFU with the same decay channels using events where the accompanying $B$ meson is fully reconstructed in hadronic modes~\cite{Belle-II:2023kkm}. 

In this work, we exploit an independent dataset to further study LFU in \bdslnu decays, and to measure the form factors, which are essential to predict $ {\cal R }(D^{(*)})$ within and beyond the SM. 

We reconstruct \bdslnu decays with the subsequent \DstDpi and \DKpi decays, without explicit reconstruction of the other $B$ meson produced in the $e^+e^- \to \Upsilon(4S) \to B^0\Bzbar$ process. The measurement uses \lumi of data recorded by the Belle~II experiment. A novel approach is developed to reconstruct the four kinematic variables that fully describe the \bdslnu decay. We report partial decay rates in intervals (bins) of these kinematic variables corrected for detector distortions and acceptance effects. 
Furthermore, we report the total rate, the values of the form factors, and \Vcb.

The remainder of this paper is organized as follows. Section~\ref{sec:theory} introduces the theoretical formalism of \bdslnu decays and the measured observables. An overview of the Belle~II subdetectors and the dataset is given in Sec.~\ref{sec:dataset}. In Secs.~\ref{sec:reconstruction} and \ref{sec:kinematic}, we summarize the event selection and the reconstruction of the four kinematic variables. Section~\ref{sec:decay rates} details the signal extraction and unfolding procedure and Sec.~\ref{sec:uncertainties} lists the systematic uncertainties affecting the measurement. The measured value of \Vcb and form-factor parameters are discussed in Sec.~\ref{sec:Vcb}. Finally, Sec.~\ref{sec:summary} presents a summary and our conclusions.

%%%%%%%%%%%%%%%%%%%%%%%%%%%%%%%%%%%

\section{Theory of \bdslnu decays}\label{sec:theory}

In the SM and the heavy-quark symmetry basis~\cite{Korner:1989qb,Richman:1995wm}, the transition between $B$ and $D^*$ mesons is represented in terms of four independent form factors $h_{A_{1-3}, V}$,
\begin{linenomath*}
\begin{align}\label{eq:hadronic current}
\frac{\langle D^*(p_{D^*})|\bar{c}\gamma^\mu b|\bar{B}(p_B)\rangle}{\sqrt{m_Bm_{D^*}}}=~&i~h_V \, \varepsilon^{\mu\nu\alpha\beta} \, \epsilon^*_\nu  \,v'_\alpha \, v_\beta\, ,\\
\frac{\langle D^*(p_{D^*})|\bar{c}\gamma^\mu\gamma_5 b|\bar{B}(p_B)\rangle}{\sqrt{m_Bm_{D^*}}}=~&h_{A_1} \,(w+1) \, \epsilon^{*\mu}-h_{A_2} \, (\epsilon^*\cdot v)\\ \nonumber
&\times v^\mu-h_{A_3} \, (\epsilon^*\cdot v) \, v'^\mu \, .
\end{align}
\end{linenomath*}
Here $v=p_B/m_B$ and $v'=p_{D^*}/m_{D^*}$ are the four-velocities of $B$ and $D^*$ mesons, respectively, with $p_B$ and $p_{D^*}$ denoting four-momenta of $B$ and $D^*$ mesons, and $m_B$ and $m_{D^*}$ denoting masses of $B$ and $D^*$ mesons, respectively. Further, $\epsilon^*$ is the polarization of the $D^*$ meson and $\varepsilon^{\mu\nu\alpha\beta}$ is the Levi-Civita tensor. The four form factors $h_{A_{1-3}, V}$ parametrize the nonperturbative physics of the $B \to D^*$ transition as functions of the recoil parameter $w = v \cdot v'$, which is related to the squared four-momentum transferred from the $B$ meson to the $D^{*}$ meson as
\begin{linenomath*}
\begin{align} 
q^2 = \left(p_B - p_{D^*} \right)^2 = m_B^2 + m_{D^*}^2 - 2 m_B \, m_{D^*} \, w  \, . 
\end{align}
\end{linenomath*}
In the light lepton mass limit, the \bdslnu decay is fully described by $h_{A_1}$ and two form-factor ratios,
\begin{linenomath*}
\begin{align}\label{eq:R1,R2}
 R_1 = \frac{h_V}{h_{A_1}} \, , \qquad R_2 = \frac{ h_{A_3} + r^* h_{A_2}}{h_{A_1}} \, ,
\end{align}
\end{linenomath*}
with $r^* = m_{D^*}/m_B$. 

An alternative parametrization is given by the helicity basis~\cite{Boyd:1997kz,Boyd:1995sq}. The three form factors in this basis, $g,f$, and ${\cal F}_1$ are related to the heavy-quark basis via
\begin{linenomath*}
\begin{align}
 f &= m_B \sqrt{r^*} (w+1) \, h_{A_1} \, , \qquad g = \frac{1}{m_B \sqrt{r^*}} \, R_1 \, h_{A_1} \, , \\
 {\cal F}_1 &= m_B^2 \, \sqrt{r^*} (w+1) \left( w - r^* - (w-1) R_2 \right) \, h_{A_1}.
\end{align}
\end{linenomath*}

\begin{figure}
    \centering
    \includegraphics[width=0.8\columnwidth]{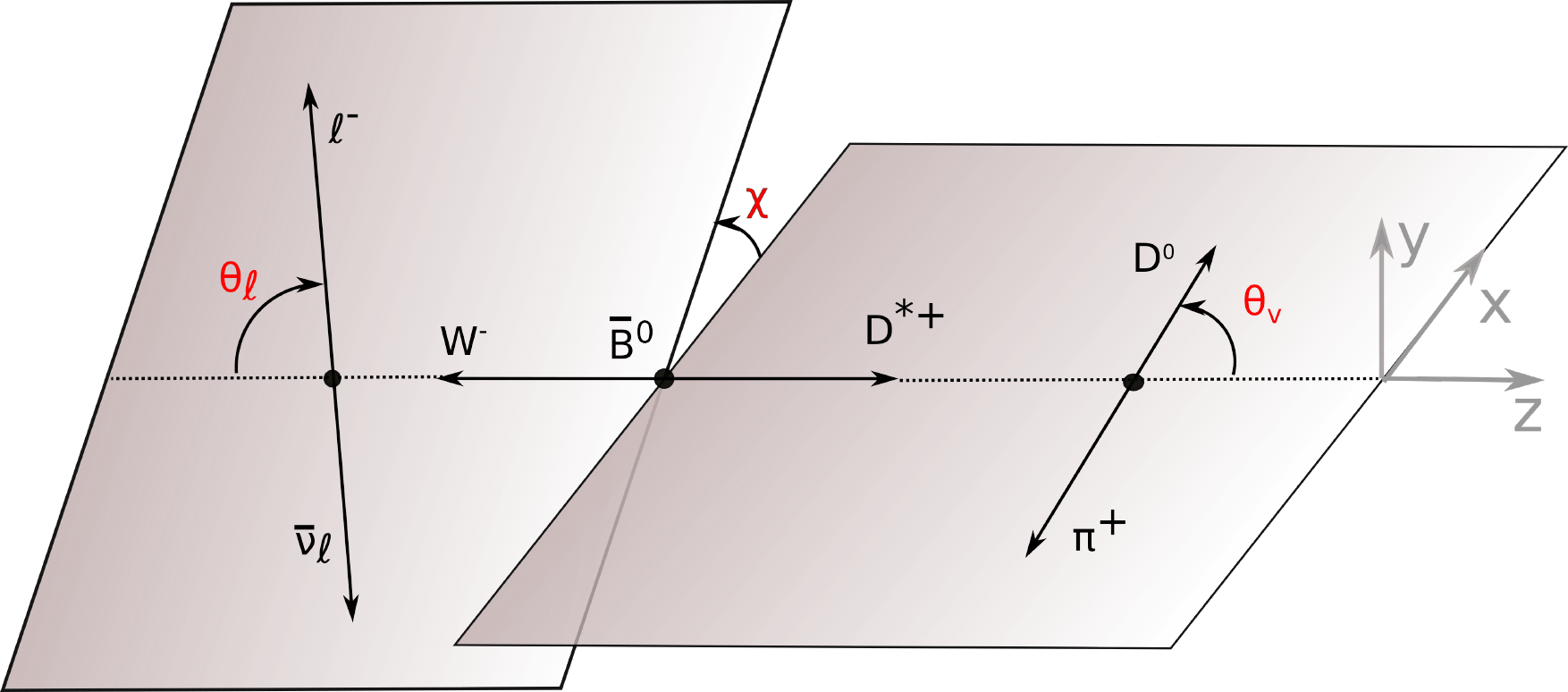}
    \caption{Sketch of the helicity angles $\theta_\ell$, $\theta_V$, and $\chi$ that characterize the \bdslnu decay.}
    \label{fig:helicity}
\end{figure}

The \bdslnu decay rate is fully parametrized by the recoil parameter $w$ and three decay angles $\theta_\ell$, $\theta_V$, and $\chi$, defined as follows (see also Fig.~\ref{fig:helicity}).
\begin{itemize}
    \item[-] The angle $\theta_\ell$ is the angle between the direction of the charged lepton and the direction  opposite to the $B$ meson in the virtual $W$ boson rest frame. 
    \item[-] The angle $\theta_V$ is the angle between the direction of the $D$ meson and the direction opposite to the $B$ meson in the $D^*$ meson rest frame. 
    \item[-] The angle $\chi$ is the azimuthal angle between the two decay planes spanned by the $W$ boson and $D^*$ meson decay products, and defined in the rest frame of the $B$ meson. 
\end{itemize}
Integrating over the angles, the \bdslnu decay rate is expressed as
\begin{align}
\frac{\mathrm{d} \Gamma(\bdslnu)}{\mathrm{d} w} & = \frac{G_F^2|V_{cb}|^2 \, \eta_{\rm EW}^{2} \,  m_B^5c^4}{48 \pi^3\hbar^7}\, \nonumber \\
  & \quad \times  (w^2-1)^{1/2}\, (w + 1)^2\, r^{*3} (1- r^*)^2 \nonumber \\
  & \quad \times \bigg[1 + \frac{4w}{w+1}\frac{1- 2 w r^* + r^{*2}}{(1 - r^*)^2} \bigg] \mathcal{F}(w)^2\,,
\end{align}
with $G_F$ and $\eta_{\rm EW} \simeq 1.0066$~\cite{Sirlin:1981ie} denoting the Fermi coupling constant and electroweak correction, respectively, and 
\begin{align}\label{eq:curly_F}
\mathcal{F}(w)^2 & = h_{A_1}^2 \bigg\{ 2(1 - 2 w r^* + r^{*2})
  \bigg(1 + R_1^2\, \frac{w-1}{w+1}\bigg)  \nonumber \\
   & \qquad\qquad\qquad  + \big[(1 - r^*) + (w-1)\big( 1 - R_2\big) \big]^2 \bigg\} \nonumber \\
	& \qquad\quad \times \bigg[(1 - r^*)^2 + \frac{4w}{w+1}\big(1- 2 wr^* + r^{*2}\big) \bigg]^{-1}\,.
\end{align}

%%%%%%%%%%%%%%%%%%%%%%%%%%%%%%
\subsection{Boyd-Grinstein-Lebed parametrization}\label{sec:BGL}

References~\cite{Boyd:1995sq,Boyd:1997kz} (Boyd, Grinstein, and Lebed; BGL) utilize dispersive bounds and expand the helicity-basis form factors with a conformal parameter,
\begin{linenomath*}
\begin{align}\label{eq:z}
 z = \frac{ \sqrt{w+1} - \sqrt{2} }{  \sqrt{w+1} + \sqrt{2} } \, ,
\end{align}
\end{linenomath*}
in terms of coefficients $\{ a_n, b_n, c_n \}$ as~\cite{Grinstein:2017nlq}
\begin{linenomath*}
\begin{align}
 g(z)&=\frac{1}{P_g(z)\phi_g(z)}\sum^{n_a-1}_{n=0}a_nz^n, \label{eq:g(z)}\\
 f(z)&=\frac{1}{P_f(z)\phi_f(z)}\sum^{n_b-1}_{n=0}b_nz^n,\\
 {\cal F}_1(z)&=\frac{1}{P_{{\cal F}_1}(z)\phi_{{\cal F}_1}(z)}\sum^{n_c-1}_{n=0}c_nz^n,\, \label{eq:F1(z)}
\end{align}
\end{linenomath*}
with $n_{a/b/c}$ the order of the expansion. Further, $P$ and $\phi$ denote the corresponding Blaschke factors and the outer functions, respectively. Note that in the expansion $b_0$ and $c_0$ are not independent quantities but related via
\begin{linenomath*}
\begin{equation}\label{eq:c0}
     c_0 = \left(\frac{\left(m_B-m_{D^*}\right)\phi_{{\cal F}_1}(0)}{\phi_f(0)}\right)b_0 \, .
\end{equation}
\end{linenomath*}

\subsection{Caprini-Lellouch-Neubert parametrization}\label{sec:CLN}

Reference~\cite{Caprini:1997mu} (Caprini, Lellouch, and Neubert; CLN) uses dispersive bounds and quark-model input to reduce the number of parameters required to describe the form factors. The  form-factor $h_{A_1}$ is expanded in the conformal parameter $z$ [Eq.~\ref{eq:z}] with a single slope parameter $\rho^2$, 
\begin{linenomath*}
\begin{align}
h_{A_1}(z) \,= & \, h_{A_1}(w=1) \bigg(1-8\rho^2z+(53\rho^2-15)z^2 \\ \nonumber
    & \qquad\qquad\qquad\qquad -(231\rho^2-91)z^3 \bigg).
\end{align}
 \end{linenomath*}    
The remaining form factors are parametrized using the ratios 
\begin{linenomath*}
\begin{align}
    R_1(w)\,=&\, R_1(1)-0.12(w-1)+0.05(w-1)^2,\\ 
    R_2(w) \, =&\, R_2(1)+0.11(w-1)-0.06(w-1)^2.
\end{align}
\end{linenomath*}
Therefore the form factors are fully parametrized by $\rho^2, R_{1}(1)$, and $R_{2}(1)$.

%%%%%%%%%%%%%%%%%%%%%%%%%%%%%%%%%%%

\section{Belle~II detector and data sample}\label{sec:dataset}

The Belle~II detector~\cite{Abe:2010sj, ref:b2tip} operates at the SuperKEKB asymmetric-energy electron-positron collider~\cite{Akai:2018mbz} and is located at the KEK laboratory. The detector consists of several nested detector subsystems arranged around the beam pipe in a cylindrical geometry. The innermost subsystem is the vertex detector, which includes two layers of silicon pixel detectors and four outer layers of silicon strip detectors. For the data used in this work, the outer pixel layer is installed in only a small part of the solid angle, while the remaining vertex-detector layers are fully instrumented. Most of the tracking volume consists of a helium and ethane-based small cell drift chamber (CDC). We define the $z$ axis parallel to the CDC axis of symmetry and directed along the boost direction. The azimuthal angle $\phi$ and the polar angle $\theta$ are defined with respect to this axis. The CDC covers a $\theta$ range between 17\degree and 150\degree and the full $\phi$ range. Outside the drift chamber, a time-of-propagation detector provides charged-particle identification in the barrel region, covering a polar angle $32.2\degree<\theta<128.7\degree$. In the forward end cap, covering a range of $12.4\degree<\theta<31.4\degree$, charged-particle identification is provided by a proximity-focusing, ring-imaging Cherenkov detector with an aerogel radiator (ARICH). Further out is the electromagnetic calorimeter (ECL) made of CsI(Tl) crystals. It consists of the barrel, forward end cap and backward end cap, which covers a range of $130.7\degree<\theta<155.1\degree$. 
An axial magnetic field is provided by a superconducting solenoid situated outside the calorimeter. Multiple layers of scintillators and resistive-plate chambers, located between the magnetic flux-return iron plates, constitute the $K^0_L$ and muon identification system (KLM). 

This analysis uses data collected between 2019 and 2021 at a c.m.\ energy of $\sqrt{s} = 10.58$~GeV and corresponding to an integrated luminosity of \lumi. The energies of the electron and positron beams are $7$ and $4 \, \mathrm{GeV}$, respectively.
The number of $B$-meson pairs is determined, using event-shape variables, to be \NBB. In addition, $12 \, \mathrm{fb}^{-1}$ of collision data recorded at a c.m.\ energy of 10.52~GeV are used to validate the modeling of continuum including $e^+e^-\to \tau^+\tau^-$ and $e^+e^-\to q\bar q$ processes, where $q$ indicates an $u$, $d$, $s$, or $c$ quark.

Simulated Monte Carlo (MC) samples of \mbox{\bdslnu} signal decays, with the subsequent \mbox{$D^{*+}\to D^0\pi^+$} and $D^0 \to K^- \pi^+$ decays, are used to obtain the template shapes for the signal extraction, and determine signal kinematic distributions and reconstruction efficiencies. These samples are generated using the \texttt{EvtGen}~\cite{Lange:2001uf} and \texttt{PYTHIA8}~\cite{Sjostrand:2007gs} computer programs with the other $B$ meson in the event allowed to decay generically. Samples of simulated background events are used to model kinematic distributions of background processes. 
These include 1~ab$^{-1}$ samples of $e^+ e^- \to  B\overline{B}$ events in which $B$ mesons decay generically, generated with \texttt{EvtGen} and \texttt{PYTHIA8}.
Samples of $e^+e^-\to q\bar q$ events are simulated with the \texttt{KKMC} generator~\cite{Jadach:1999vf} interfaced with \texttt{PYTHIA8}.
Further, $e^+e^-\to \tau^+\tau^-$ events are simulated with \texttt{KKMC}, and interfaced with \texttt{TAUOLA}~\cite{Jadach:1990mz}.
Interactions of detectors and particles are simulated by \texttt{GEANT4}~\cite{GEANT4:2002zbu}. All recorded data and simulated samples are processed and analyzed using the Belle~II software~\cite{Kuhr:2018lps,the_belle_ii_collaboration_2021_5574116}.

%%%%%%%%%%%%%%%%%%%%%%%%%%%%%%%%%%%

\section{Event selection}\label{sec:reconstruction}

We select \bdslnu decays with which we determine the distributions of $w$, $\cos\theta_\ell$, $\cos\theta_V$, and $\chi$ as well as the absolute branching fraction.
 
To suppress beam backgrounds, charged particles are required to have a distance of closest approach to the interaction point of less than 4.0 cm along the $z$ direction and less than 2.0 cm in the transverse $r-\phi$ plane. A summary of the reconstruction algorithms for trajectories of charged-particles (tracks) used in this analysis is given in Ref.~\cite{BelleIITrackingGroup:2020hpx}. We require all tracks to be within the angular acceptance of the CDC. The momenta of electron and muon candidates in the c.m.\ frame are required to be larger than 1.2 GeV/$c$ and smaller than 2.4 GeV/$c$. Further, electron candidates are selected using particle identification (ID) likelihoods based on CDC, ARICH, ECL, and KLM information. 
Information from the time-of-propagation detector is included in addition to information from the CDC, ARICH, ECL, and KLM to identify muon candidates. 
The efficiencies to identify electrons and muons are 88\% and 91\%, respectively. The misidentifiation rates of hadrons, including pions and kaons, as electrons and muons are 0.2\% and 3\%, respectively. 
The efficiencies for electron and muon identification are calibrated using $J/\psi\to\ell^+\ell^-$, $e^+e^-\to\ell^+\ell^-(\gamma)$, and $e^+e^-\to(e^+e^-)\ell^+\ell^-$ channels. The misidentification probabilities of charged kaons as leptons are calibrated using the $D^{*+}\rightarrow D^0 (\rightarrow K^- \pi^+) \pi^+$ process. The rate of misidentified charged pions is studied using $K_S^0\rightarrow \pi^+\pi^-$ and $e^+e^-\rightarrow \tau^+(1\text{-prong})\tau^-(3\text{-prong})$ decays. 

Neutral $D$ candidates are reconstructed from charged kaon and pion candidates and their invariant masses are required to be within $ 15$ MeV/$c^2$ from the known $D^0$ mass, corresponding to a range of $\pm 3.4$ times the mass resolution.
To reconstruct $D^{*+}$ candidates, the $D^0$ candidates are combined with low-momentum pion candidates (slow pions) selected from the remaining charged particles with momentum below 0.4~GeV/$c$. To reduce the fraction of incorrectly reconstructed $D^{*+}$ candidates, the mass difference of $D^{*+}$ and $D^0$ candidates $\Delta M=M(K\pi\pi)-M(K\pi)$ is required to be in the range $[0.141, 0.156]$~GeV/$c^2$, with correctly reconstructed $D^{*+}$ candidates peaking at a value of $m_{D^{*+}}=2.010$~GeV/$c^2$, with a resolution of 5~MeV/$c^2$.

Continuum is suppressed by requiring that the ratio between the second- and the zeroth-order Fox-Wolfram moments~\cite{PhysRevLett.41.1581} is less than $0.3$. A limit on the $D^{*+}$ momentum in the c.m.\ frame, $p_{D^{*+}}^{\mathrm{c.m.}}<2.5$ GeV/$c$, further rejects $D^{*+}$ candidates from $e^+ e^- \to c\bar c$ events. The sum of the reconstructed energy in the c.m.\ frame is required to be larger than 4~GeV. All of the selection criteria together result in efficiencies of $22.0\%$ and $23.5\%$ for the \bdsenu and \bdsmunu decays, respectively. We find $1.06$ candidates per event, which is in good agreement with the candidate multiplicity of the simulated samples. We retain all candidates per event, cf. Ref.~\cite{Koppenburg:2017zsh}. The event selection strategy is optimized and tested using simulated samples and no bias on the selection efficiency is observed. 

\section{Reconstruction of kinematic observables }\label{sec:kinematic}

\begin{figure*}
    \centering
    \includegraphics[width=0.88\textwidth]{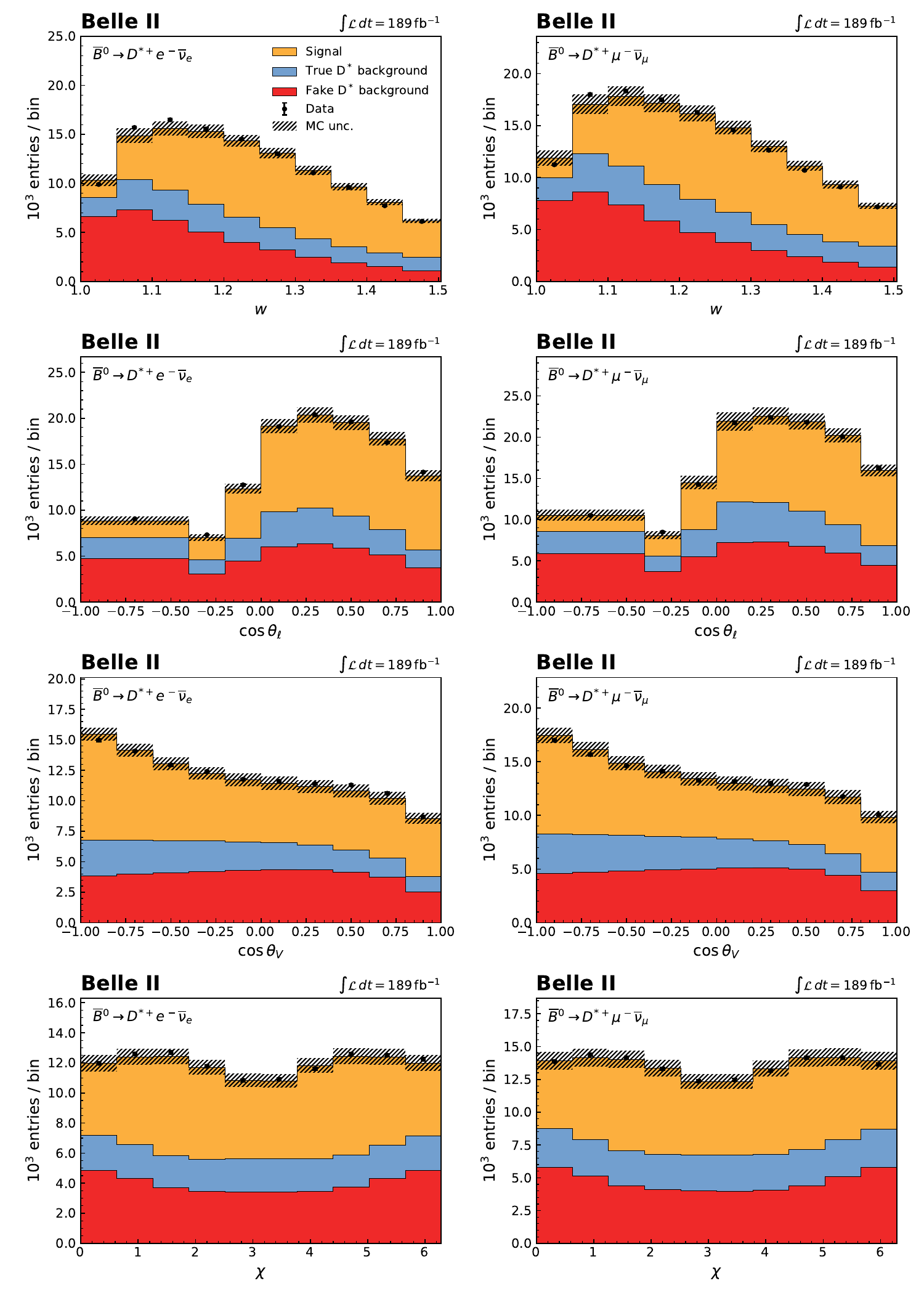}
    \caption{Distributions of observed kinematic variables \kinematic for \bdsenu (left) and \bdsmunu (right) candidates reconstructed in data with expected distributions from simulation overlaid. In all panels, simulated samples are shown separately for signal, true $D^*$ background, and fake $D^*$ background and weighted according to luminosities. The hatched area represents the statistical uncertainty due to the finite size of the simulated samples, and systematic uncertainty arising from the lepton identification, slow pion reconstruction, and tracking efficiency of $K$, $\pi$, and $\ell$.
    } \label{fig:kinematic variable}
\end{figure*}

The signal $B$-meson energy and magnitude of the momentum in the c.m.\ frame is inferred from the beam energy $E^{\mathrm{c.m.}}_{\mathrm{Beam}} = \sqrt{s}/2$,
\begin{linenomath*}
\begin{equation}
    E^{\mathrm{c.m.}}_B=E^{\mathrm{c.m.}}_{\mathrm{Beam}}, \quad
    |\vec{p}^{\,\mathrm{c.m.}}_B|=\sqrt{(E^{\mathrm{c.m.}}_{\mathrm{Beam}})^2-m^2_{B}c^4}/c \, .
\end{equation}
\end{linenomath*}
Here $m_B = \SI{5.279}{\GeV/}c^2$ is the $B$-meson mass. With this information, we reconstruct the cosine of the angle between the $B$ meson and the $D^*\ell$ system (denoted by $Y$) via
\begin{linenomath*}
\begin{equation}
    \cos\theta_{BY}=\frac{2E^{\text{c.m.}}_BE^{\text{c.m.}}_Y-m^2_Bc^4-m^2_Yc^4}{2|\vec{p}^{\,\text{c.m.}}_B||\vec{p}^{\,\text{c.m.}}_Y|c^2},
\end{equation}
\end{linenomath*}
where $E^{\text{c.m.}}_Y$, $|\vec{p}^{\,\text{c.m.}}_Y|$, and $m_Y$ are the energy, magnitude of momentum, and mass of the reconstructed $D^*\ell$ system, respectively.  If a $D^* \ell$ pair does not originate from a \bdslnu\ decay, then $|\cos\theta_{BY}|$ can exceed unity. One such possible background are $B \to D^{**} \ell \bar \nu_\ell$ decays with $D^{**}$ denoting $D_0^*, D_1', D_1, D_2^*$ final states, where missing final state particles may shift $\cos\theta_{BY}$ to large negative values. We retain all candidates in the range $\cos\theta_{BY} \in [-4, 2]$ to separate signal decays and background processes. 

To reconstruct the recoil parameter $w$ and three helicity angles, the direction of the signal $B$ momentum is needed. It is constrained to lie on a cone around the direction of the $Y$ system with an opening angle $2\theta_{BY}$. To constrain the position of the $B$ momentum on the cone, we exploit the polarization of the $\Upsilon(4S)$, which produces $B$ pairs with an angular distribution of $\sin^2\theta$ with respect to the beam axis. This method was introduced by the BaBar experiment in Refs.~\cite{Gill:2004hy,BaBar:2006taf}.

Furthermore, we use the tracks and neutral energy depositions (clusters) not associated to the $Y$ system~\cite{Belle:2018ezy}. They originate from the companion $B$ meson, with additional contributions from beam backgrounds. We refer to the collection of these tracks and clusters as the rest of the event (ROE).
The ROE can be used to estimate the signal $B$ direction, as it is opposite to the direction of the ROE momentum. However, missing particles, such as  neutrinos and $K_L^0$ mesons, and resolution effects affect the ROE direction. To use both the $B$ meson angular distribution and ROE information, a novel method is developed in this work. We sample 10 $B$-meson directions that are equally spaced on the lateral surface of the cone spanned by $2\theta_{BY}$, and reconstruct $w, \cos \theta_\ell$, and $\cos \theta_V$ by calculating a weighted average, where weights are written as 
\begin{linenomath*}
\begin{equation}
    \alpha_i = (1-\hat{p}_{\mathrm{ROE}}\cdot \hat{p}_{B\, i})\sin^2\theta_{B\,i}.
\end{equation}
\end{linenomath*}
Here, $\hat{p}_{\mathrm{ROE}}$ and $\hat{p}_{B\, i}$ denote the unit vector of the ROE and one of sampled $B$ momenta, respectively, and $\theta_{B\, i}$ is the corresponding polar angle of the $B$ meson. 
However, applying the combined methodology, we observe poor resolution near $\chi = 0$ and $2 \pi$. Consequently, when reconstructing the angle $\chi$, we identify the direction on the lateral surface of the cone that minimizes the angle with respect to the vector $-\hat{p}_{\mathrm{ROE}}$ as the direction of the $B$ momentum.
The biases and resolutions on $w$ and the decay angles are listed in Table~\ref{tab:res_bias}, where biases are calculated as the medians of differences between reconstructed values and true values, and resolutions are defined as the symmetrical 68\% percentiles around the medians of residuals of reconstructed and generated values. 
The resolution of $\cos \theta_V$ is improved by $2\%$ and $7\%$ compared to using only the $B$-meson angular distribution or only ROE information, respectively. Further, the resolutions of the new method for $w$ and $\cos \theta_\ell$ are improved by $7\%$ to $12\%$ compared to previous methods~\cite{BaBar:2006taf, Belle:2018ezy}.

\begin{table}
\caption{Biases and resolutions of the reconstructed recoil parameters and decay angles.}
 \label{tab:res_bias}
 \begin{tabular}{lS[table-format=-1.4, parse-numbers=false]cc} 
  \hline  \hline
  Variable & {Bias} & Resolution \\ \hline
   $w$                & 0.001 & $0.04$ \\
   $\cos \theta_\ell$ & -0.005 & $0.10$ \\
   $\cos \theta_V$    & 0.004 &  $0.13$ \\
   $\chi$ [rad]             & 0.0004 & $0.58$ \\
    \hline \hline
\end{tabular}
\end{table}

Figure~\ref{fig:kinematic variable} shows distributions of the reconstructed kinematic variables, where simulated samples are classified as follows
\begin{itemize}
    \item Signal: the entire decay chain is reconstructed correctly. 
    \item ``True $D^*$'' background: the $D^*$ candidate is reconstructed correctly, but the $D^*\ell$ system is incorrectly reconstructed due to the misidentification of the lepton candidate or an incorrect combination of $D^*$ and $\ell$ candidates.
   This arises from continuum, $B$-meson background, or signal processes. 
    \item ``Fake $D^*$'' background: the $D^*$ candidate is misreconstructed, arising from continuum, $B$-meson background, or signal processes. 
\end{itemize}
We choose a uniform binning for the recoil parameter and the decay angles with ten bins, with the exception of $\cos \theta_\ell$, for which we choose eight bins. In general, we observe a fair agreement between simulated samples and experimental data.
For the measurement, we determine the signal and background fractions in each bin using a fit to \cosBY and $\Delta M$ distributions. 

%%%%%%%%%%%%%%%%%%%%%%%%%%%%%%%%%%%

\section{Measurement of partial decay rates}\label{sec:decay rates}

The measurement of partial decay rates in bins of the recoil parameter $w$ and the decay angles is carried out in two steps. First, backgrounds are determined using a likelihood fit and subtracted in each bin independently. The effects of the reconstruction resolution, efficiency, and detector acceptance are then corrected. The statistical correlations between the projected kinematic variables are determined. To validate the fit and unfolding procedure we generated ensembles
of simplified simulated experiments. The fits to these ensembles show no biases in central values and no under- or overcoverage of the quoted confidence intervals. 

\subsection{Signal extraction}\label{sec:signal extraction}

\begin{figure*}
    \centering
    \includegraphics[width=0.9\textwidth]{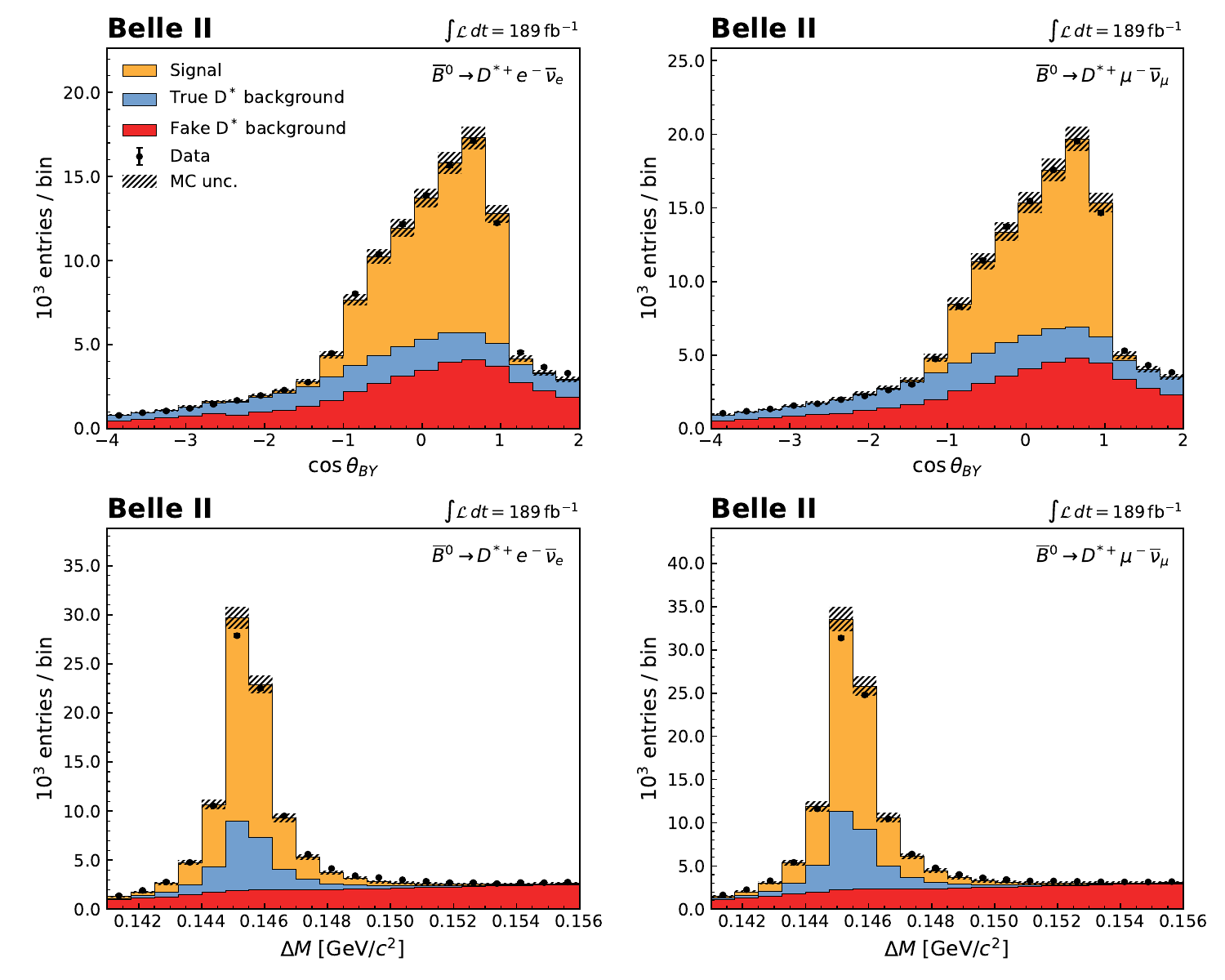}
    \caption{Distributions of reconstructed \cosBY and \DeltaM for \bdsenu (left) and \bdsmunu (right) candidates in data with expectations from simulation overlaid. The simulated samples are weighted according to integrated luminosity. The hatched area represents the uncertainty due to the finite size of the simulated sample, and systematic uncertainties arising from the lepton identification, slow-pion reconstruction, and tracking efficiency of the $K$, $\pi$, and $\ell$.
    }\label{fig:cosBY&DelM}
\end{figure*}

The signal yield in each bin of the kinematic variables is extracted using two-dimensional likelihood fits to the binned \cosBY and \DeltaM distributions, as shown in Fig.~\ref{fig:cosBY&DelM} for the electron and muon final states. The free parameters of the fits are the signal yields, the yields of the true $D^*$ background, and the yields of the fake $D^*$ background. We use simulated events for the templates. The \DeltaM fit separates signal and true $D^*$ events, which peak near $0.145$~GeV$/c^2$, from the fake $D^*$ component. 

We choose the following bin granularity with 16 bins in two dimensions: four bins in \cosBY spanning $[-4.0, -2.5,\, -1.0,\, 1.0,\, 2.0]$ and four equidistant \DeltaM bins spanning $[0.141,\, 0.156]$ GeV$/c^2$.

The expected number of events in bin $i$ of \cosBY and \DeltaM is 
\begin{align} \label{eq:frac_fit}
    \nu_i^{\text{exp}}(\boldsymbol{\nu},\boldsymbol{\theta})=\sum_k \, \nu_k \,  f_{ik}^{\mathrm{MC}}(\boldsymbol{\theta})\, ,
\end{align}
where $\nu_k$ denotes the yield for the event category $k$, which can be signal, true $D^*$ background, and fake $D^*$ background. Further, $f_{ik}^{\mathrm{MC}}$ denotes the fraction of events in bin $i$, and is written as
\begin{align} 
f^{\text{MC}}_{ik}(\boldsymbol{\theta})=\frac{p^{\text{MC}}_{ik}(1+\epsilon_{ik}\theta_{ik})}{\sum_jp^{\text{MC}}_{jk}(1+\epsilon_{jk}\theta_{jk})}\, ,
\end{align}
where $p^{\text{MC}}_{ik}$ is the probability that a $k$ category event is found in bin $i$ as determined from the simulation. This allows the shape of the template to vary according to the nuisance parameter $\theta_{jk}$ and $1\sigma$ deviation $\epsilon_{ik}$ due to the limited size of simulated samples and other systematic sources (see Sec.~\ref{sec:uncertainties}).

The likelihood function for a given bin of the recoil parameter or one of the
decay angles is
\begin{align}
    -2\ln{\cal L}(\boldsymbol{\nu}, \boldsymbol{\theta}) = -2\ln\prod_i{\cal P}(\nu_i^{\text{obs}},\, \nu_i^{\text{exp}})+ \boldsymbol{\theta}^T C^{-1}_\theta \boldsymbol{\theta},
\end{align}
and we minimize it numerically using the \texttt{iminuit} package~\cite{iminuit,James:1975dr}. Here, $\nu_i^{\text{obs}}$ is the number of observed events in data in a given bin and ${\cal P}$ denotes the Poisson distribution. Further, $C_\theta$ is the correlation matrix of the nuisance parameters. The resulting signal yields in bins of kinematic variables are provided in Appendix~\ref{app:yields}. 

\subsection{Unfolding of fitted yields}\label{sec:unfolding}

The resolution and limited acceptance of the Belle~II detector distort the kinematic variables. In order to compare them to the theory expressions of Sec.~\ref{sec:theory}, we correct the extracted number of signal events for migrations, efficiencies, and acceptance effects. The migration between observed and true values is expressed as a conditional probability of events being observed in a bin $x$ of the recoil parameter or a decay angle, given that its true value is in bin $y$,
\begin{align}
 \mathcal{M}_{xy} = P( \text{observed in bin} \,\, x | \text{true value in bin} \,\, y) \, .
\end{align}
The matrices summarizing these conditional probabilities for the electron final state are shown in Fig.~\ref{fig:migration e}, and those for the muon final state are given in Appendix~\ref{app:migration}. To correct migrations across bins, we apply the singular-value-decomposition unfolding method of Ref.~\cite{Hocker:1995kb}. The method employs a regularization parameter $k$, which dampens statistical fluctuations in the unfolded distributions. To optimize the value of $k$ for each kinematic variable, the simulated signal samples are weighted using the form-factor parameters and their $3\sigma$ uncertainties of Ref.~\cite{Ferlewicz:2020lxm}. The bias is defined as the difference of the unfolded spectrum, which is determined with the nominal migration matrix, and the underlying true spectrum. 
The $k$ values are chosen such that the bias is small, and the ratio of bias and unfolding uncertainty is low and stable. We choose $k = 7, 6, 6$, and $7$ for \kinematic, respectively. These $k$ values provide biases of similar or smaller sizes than unfolding via the inversion of the migration matrices. 

The stability of the reported results are further tested by choosing $k$ values equal to the number of bins, resulting in a less constrained curvature regularization, and by using matrix-inversion unfolding. We observe negligible shifts in the form factors and \Vcb. Their uncertainties are reported in Sec.~\ref{sec:Vcb}.

\begin{figure*}
    \centering
    \subfigure{\includegraphics[width=0.35\textwidth]{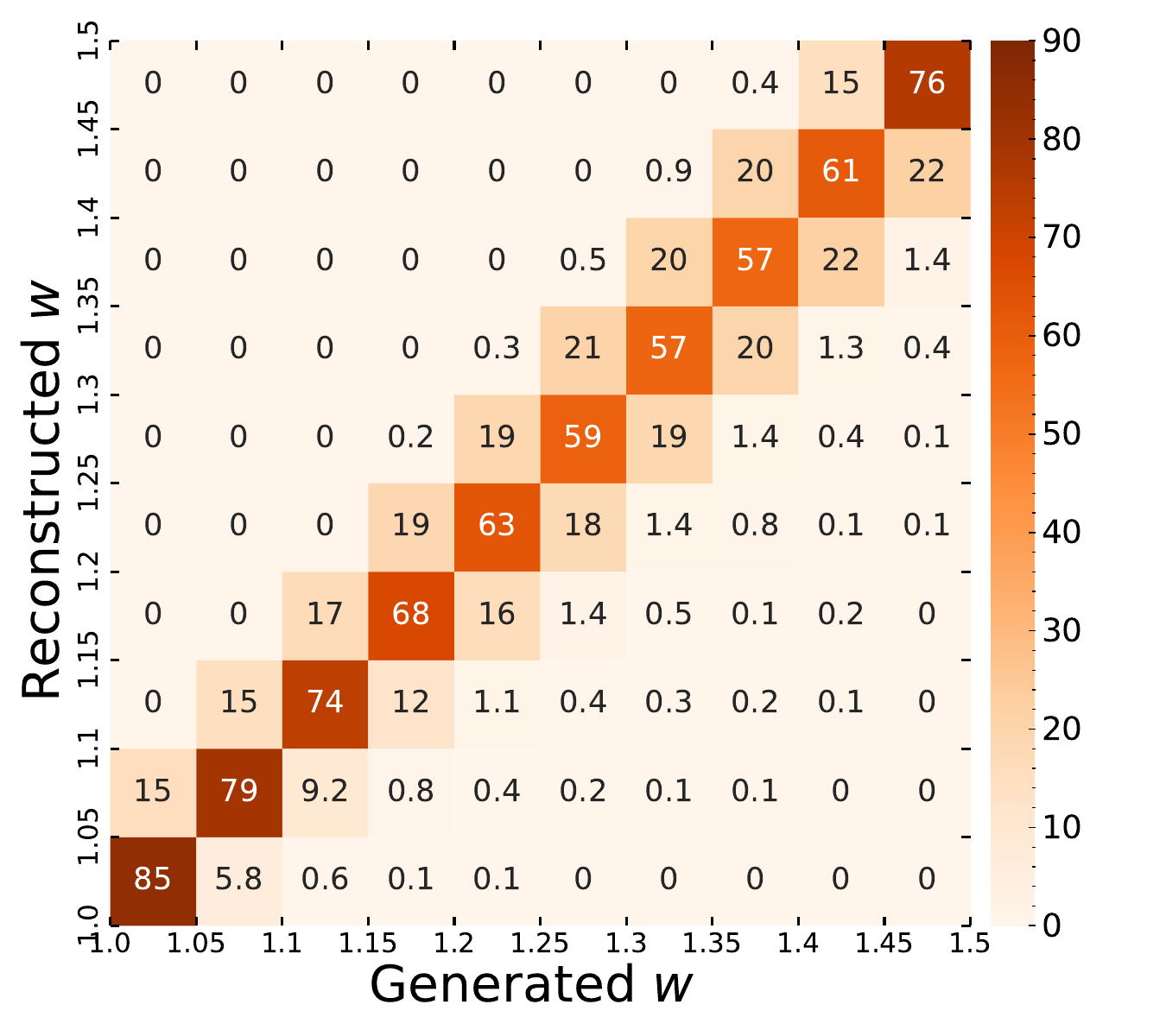}}
    \subfigure{\includegraphics[width=0.35\textwidth]{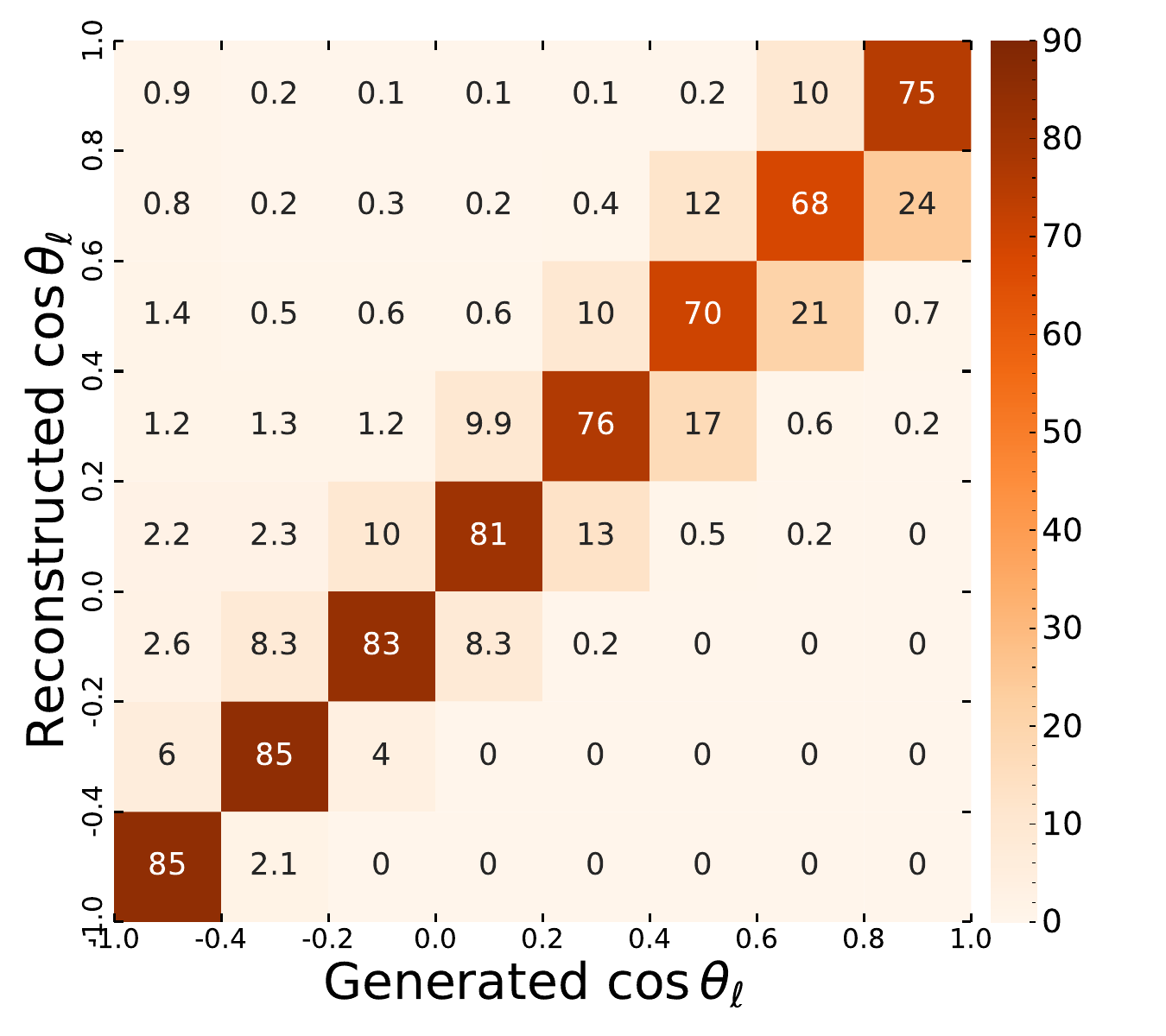}}
    \\   
    \subfigure{\includegraphics[width=0.35\textwidth]{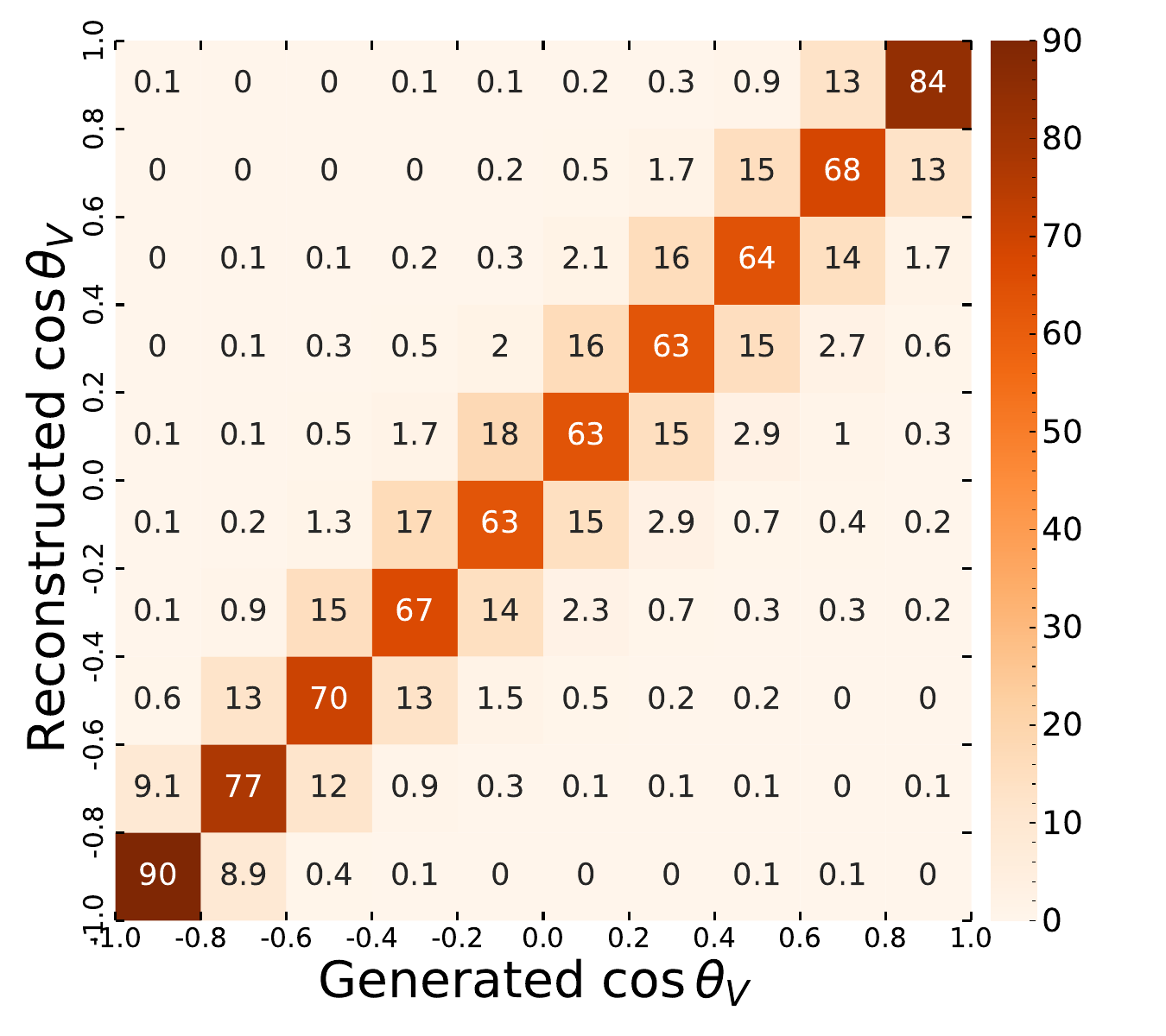}}
    \subfigure{\includegraphics[width=0.35\textwidth]{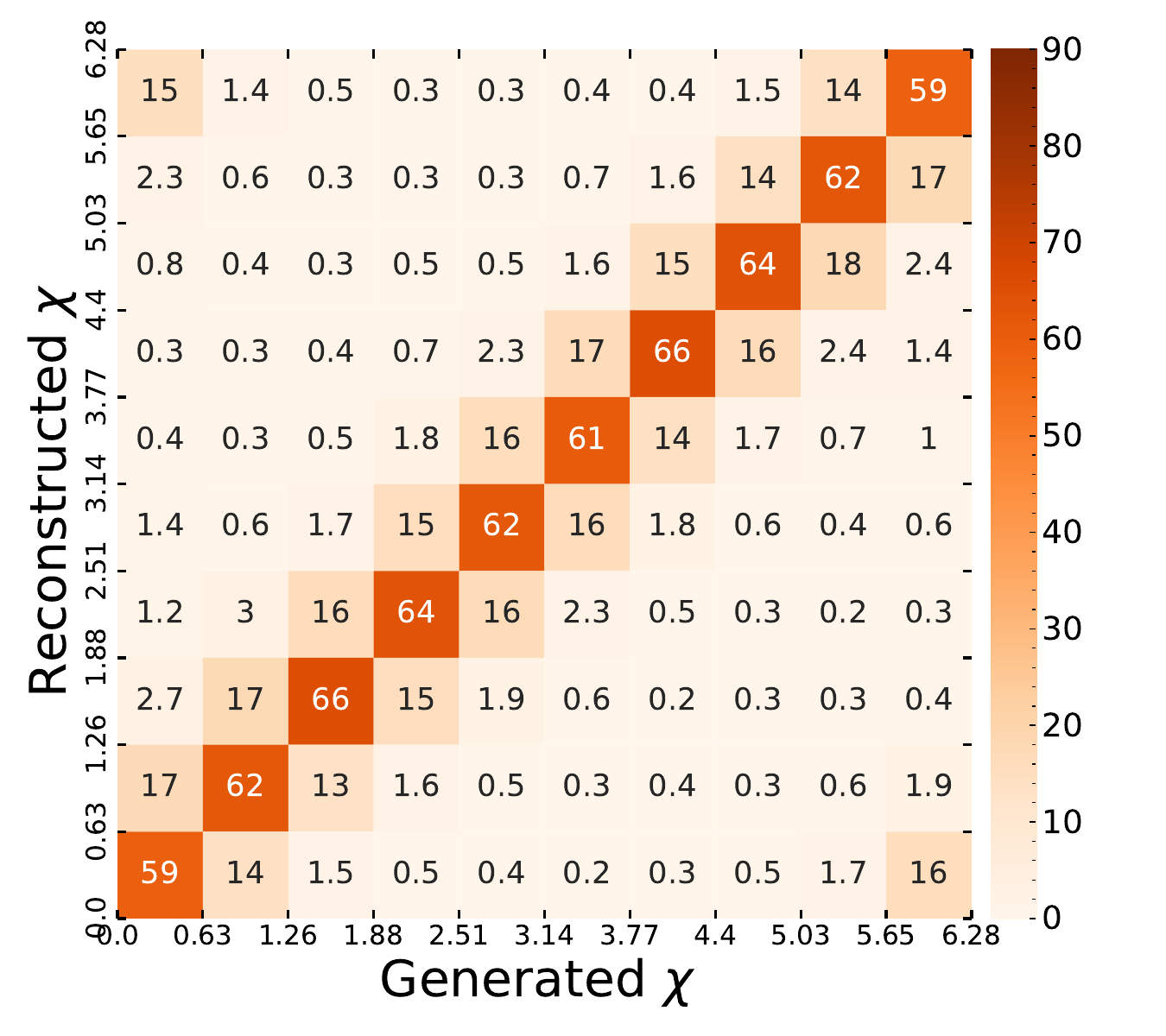}} 
    \caption{Migration matrices of the reconstructed kinematic variables in the \bdsenu decay. The values in the figures are in $10^{-2}$ units.} \label{fig:migration e}
\end{figure*}

We determine the partial decay rate in a given kinematic bin $x$ using the unfolded yields $\nu^{\text{unfolded}}_{x}$ via 
\begin{linenomath*}
\begin{equation}\label{eq:partial rate}
\Delta\Gamma_x=\frac{\nu^{\text{unfolded}}_{x}\, \hbar}{\epsilon_x \, N_{B^0} \, {\cal B}(D^{*+}\to D^0\pi^+) \, {\cal B}(D^{0}\to K^-\pi^+) \,\tau_{B^0} },
\end{equation}
\end{linenomath*}
with $\epsilon_x$ denoting the reconstruction efficiency, and $\tau_{B^0}$ denoting the $B^0$ meson lifetime~\cite{Workman:2022ynf}. The number of produced $B^0$ mesons $N_{B^0}$ is further discussed in Sec. \ref{sec:uncertainties}. The resulting partial decay rates and uncertainties are listed in Table~\ref{tab:partial rates}. The numerical values and full covariance matrices of the measured partial decay rates are available on HEPData~\cite{hepdata}.

\subsection{Statistical correlations}

To analyze the measured partial decay rates simultaneously, we determine the full statistical correlation of the four measured projections. This is done using a bootstrapping approach. We generate $10,000$ replicas of the data sample by resampling with replacement~\cite{Hayes:1988xc}. The total number of sampled events in each replica is varied according to the statistical uncertainty of the full dataset. Each replica is analyzed using the full analysis procedure (background subtraction, unfolding). The resulting statistical correlations are provided in Appendix~\ref{app:correlation}. The majority of the bins of different kinematic variables are weakly and positively correlated. Bins of the same variable may exhibit negative or positive correlations due to the unfolding procedure.

\begin{table*}
\centering
\renewcommand\arraystretch{1.2}
\caption{Measured partial decay rates $\Delta\Gamma$ (in units of $10^{-15}$ GeV/$\hbar$) and average of normalized partial decay rates $\Delta\Gamma/\Gamma$ over \bdsenu and \bdsmunu decays in bins of kinematic variables. The normalized partial decay rate in the last bin of each projection is excluded in the \Vcb determination to subtract the redundant degrees of freedom. The full (statistical and systematic) uncertainties are provided.}
\label{tab:partial rates}
\begin{tabular}{clS[table-format=1.3(1)]S[table-format=1.3(1)]S[table-format=1.3(1)]}
\hline\hline
         \multirow{2}{*}{Variable} & \multicolumn{1}{c}{\multirow{2}{*}{Bin}} & \multicolumn{2}{c}{$\Delta\Gamma$} & \multicolumn{1}{c}{$\Delta\Gamma/\Gamma$ average (in \%)} \\
 
                                    &                                          &  {~\bdsenu~} &  {~\bdsmunu~} &                        {\bdslnu} \\ \hline
              \multirow{10}{*}{$w$} &                             $[1.00, 1.05)$ &     1.34\pm 0.10     &                   1.30 \pm 0.09 &                                        6.19 \pm 0.21 \\
                                    &                             $[1.05, 1.10)$ &                  2.08\pm0.12 &                   2.11\pm0.12 &                                        9.86\pm0.22 \\
                                    &                             $[1.10, 1.15)$ &                  2.40\pm0.13 &                   2.45\pm0.13 &                                       11.39\pm0.20 \\
                                    &                             $[1.15, 1.20)$ &                  2.61\pm0.14 &                   2.60\pm0.14 &                                       12.18\pm0.19 \\
                                    &                             $[1.20, 1.25)$ &                  2.60\pm0.13 &                   2.60\pm0.13 &                                       12.16\pm0.17 \\
                                    &                             $[1.25, 1.30)$ &                  2.49\pm0.12 &                   2.43\pm0.12 &                                       11.50\pm0.17 \\
                                    &                             $[1.30, 1.35)$ &                  2.30\pm0.11 &                   2.29\pm0.11 &                                       10.72\pm0.17 \\
                                    &                             $[1.35, 1.40)$ &                  2.07\pm0.10 &                   2.07\pm0.10 &                                        9.67\pm0.18 \\
                                    &                             $[1.40, 1.45)$ &                  1.83\pm0.09 &                   1.80\pm0.09 &                                        8.47\pm0.17 \\
                                    &                             $[1.45, 1.51)$ &                  1.67\pm0.09 &                   1.70\pm0.10 &                                                {} \\ \hline
 \multirow{8}{*}{$\cos\theta_\ell$} &                           $[-1.00, -0.40)$ &                  3.89\pm0.33 &                   4.10\pm0.39 &                                       18.94\pm0.79 \\
                                    &                           $[-0.40, -0.20)$ &                  2.00\pm0.14 &                   2.07\pm0.16 &                                        9.60\pm0.27 \\
                                    &                            $[-0.20, 0.00)$ &                  2.28\pm0.12 &                   2.26\pm0.14 &                                       10.63\pm0.19 \\
                                    &                             $[0.00, 0.20)$ &                  2.51\pm0.12 &                   2.56\pm0.14 &                                       11.86\pm0.24 \\
                                    &                             $[0.20, 0.40)$ &                  2.73\pm0.13 &                   2.63\pm0.13 &                                       12.54\pm0.25 \\
                                    &                             $[0.40, 0.60)$ &                  2.70\pm0.13 &                   2.70\pm0.13 &                                       12.68\pm0.24 \\
                                    &                             $[0.60, 0.80)$ &                  2.54\pm0.12 &                   2.57\pm0.12 &                                       12.01\pm0.24 \\
                                    &                             $[0.80, 1.00)$ &                  2.52\pm0.12 &                   2.49\pm0.12 &                                                {} \\ \hline
   \multirow{10}{*}{$\cos\theta_V$} &                           $[-1.00, -0.80)$ &                  2.89\pm0.13 &                   3.02\pm0.14 &                                       13.86\pm0.27 \\
                                    &                           $[-0.80, -0.60)$ &                  2.38\pm0.10 &                   2.32\pm0.11 &                                       11.00\pm0.18 \\
                                    &                           $[-0.60, -0.40)$ &                  1.98\pm0.09 &                   1.93\pm0.09 &                                        9.14\pm0.13 \\
                                    &                           $[-0.40, -0.20)$ &                  1.67\pm0.08 &                   1.65\pm0.08 &                                        7.75\pm0.11 \\
                                    &                            $[-0.20, 0.00)$ &                  1.54\pm0.08 &                   1.53\pm0.08 &                                        7.18\pm0.10 \\
                                    &                             $[0.00, 0.20)$ &                  1.56\pm0.08 &                   1.58\pm0.09 &                                        7.37\pm0.11 \\
                                    &                             $[0.20, 0.40)$ &                  1.73\pm0.09 &                   1.77\pm0.10 &                                        8.20\pm0.12 \\
                                    &                             $[0.40, 0.60)$ &                  2.05\pm0.11 &                   2.04\pm0.11 &                                        9.59\pm0.14 \\
                                    &                             $[0.60, 0.80)$ &                  2.48\pm0.13 &                   2.42\pm0.14 &                                       11.48\pm0.17 \\
                                    &                             $[0.80, 1.00)$ &                  3.07\pm0.17 &                   3.09\pm0.18 &                                                {} \\ \hline
           \multirow{10}{*}{$\chi$} &                             $[0.00, 0.63)$ &                  1.82\pm0.11 &                   1.85\pm0.11 &                                        8.59\pm0.21 \\
                                    &                             $[0.63, 1.26)$ &                  2.20\pm0.11 &                   2.24\pm0.12 &                                       10.42\pm0.16 \\
                                    &                             $[1.26, 1.88)$ &                  2.55\pm0.13 &                   2.50\pm0.13 &                                       11.82\pm0.18 \\
                                    &                             $[1.88, 2.51)$ &                  2.24\pm0.11 &                   2.24\pm0.11 &                                       10.51\pm0.16 \\
                                    &                             $[2.51, 3.14)$ &                  1.85\pm0.09 &                   1.83\pm0.10 &                                        8.62\pm0.15 \\
                                    &                             $[3.14, 3.77)$ &                  1.89\pm0.10 &                   1.85\pm0.10 &                                        8.75\pm0.14 \\
                                    &                             $[3.77, 4.40)$ &                  2.19\pm0.11 &                   2.21\pm0.11 &                                       10.31\pm0.16 \\
                                    &                             $[4.40, 5.03)$ &                  2.47\pm0.12 &                   2.56\pm0.13 &                                       11.82\pm0.17 \\
                                    &                             $[5.03, 5.65)$ &                  2.24\pm0.11 &                   2.30\pm0.12 &                                       10.67\pm0.15 \\
                                    &                             $[5.65, 6.28)$ &                  1.88\pm0.11 &                   1.75\pm0.10 &                                                {} \\
\hline\hline
\end{tabular}
\end{table*}

%%%%%%%%%%%%%%%%%%%%%%%%%%%%%%%%%%%

\section{Systematic uncertainties}\label{sec:uncertainties}

Several systematic uncertainties affect the measured partial rates. They are grouped into uncertainties stemming from the background subtraction and uncertainties affecting the unfolding procedure and the efficiency corrections. A detailed breakdown for each measured partial decay rate is given in Appendix~\ref{app:uncertainty}.  
The statistical and systematic correlation matrices are given in Appendix~\ref{app:correlation}.

\subsection{Background subtraction}

The background subtraction is sensitive to the signal and background template shapes in \cosBY and \DeltaM. To validate the modeling, we reconstruct a sample of same-sign $D^{*+} \ell^+$ events, which are free of our signal decay. We observe a fair agreement in the analyzed range of \cosBY, but observe some deviations of data from simulation for ${\cosBY > 2.5}$. We derive correction factors for both background templates based on  $D^{*+} \ell^+$ events. We use the high \DeltaM region to derive a correction factor for fake $D^*$ contributions, and the region near \DeltaM of 0.145 GeV/$c^2$ to determine a correction for the true $D^*$ contribution. The resulting correction factors are in the range [0.85, 1.15]. The full difference between applying and not applying this correction is taken as the systematic uncertainty from background modeling. The systematic uncertainties from the correction factors are treated as uncorrelated. The impact of varying the assumed $B \to D^{**} \ell \bar \nu_\ell$ background composition is also studied and is found to be negligible.

\subsection{Size of simulated samples}

We propagate the statistical uncertainty from the limited size of the simulated sample into the signal and background shapes, migration matrices, and signal efficiencies. For the signal and background shapes, we use nuisance parameters to allow the template shapes in \cosBY and \DeltaM to vary within their statistical uncertainties. The uncertainties associated with the finite size of simulated samples are uncorrelated given that they are determined independently bin by bin.

\subsection{Lepton identification}

We use bin-wise correction factors as functions of the laboratory momentum and polar angle of the lepton candidates. To determine the uncertainties, we produce 400 replicas of the bin-wise correction factors that fluctuate each factor within its uncorrelated statistical uncertainty and its correlated systematic uncertainty. For each replica, the migration matrices and efficiencies are redetermined and the signal extraction is repeated. 

\subsection{Tracking efficiency}

We assign a track selection uncertainty of 0.3\% per track on kaon, pion, and lepton tracks due to imperfect knowledge of the track-selection efficiency. This uncertainty is determined using a control sample of $e^+ e^- \to \tau^+ \tau^-$ events, and is assumed to be fully correlated between all tracks. 

\subsection{Slow-pion reconstruction efficiency}

The slow-pion efficiency is determined using $B^0 \to D^{*-} \pi^+$ decays and calculated relative to the tracking efficiency at momenta larger than 200 MeV/$c$ in the laboratory frame. We derive correction factors for three momentum bins spanning $[0.05, 0.12, 0.16, 0.20] \, \mathrm{GeV}/c$. To propagate the impact on the measurement, we produce 400 replicas of the correction weights, taking into account correlations. For each replica, migration matrices and efficiencies are redetermined, and the partial decay rates are remeasured. 

\subsection{Number of $B^0$ mesons}
The number of $B\Bbar$ pairs, \NBB, is used to determine the total number of $B^0$ mesons in the dataset,
\begin{align}
N_{B^0} = 2  N_{B{\kern 0.18em\overline{\kern -0.18em B}}} \,  (1+f_{+0})^{-1},
\end{align}
with $f_{+0}={\cal B}(\Upsilon(4S)\rightarrow B^+B^-)/{\cal B}(\Upsilon(4S)\rightarrow B^0{\kern 0.18em\overline{\kern -0.18em B}}^0)=1.065 \pm 0.052$ \cite{Belle:2022hka}. The uncertainties from both $N_{B {\kern 0.18em\overline{\kern -0.18em B}}}$ and $f_{+0}$ are propagated into the measured partial decay rates. 

\subsection{External inputs}

In Eq.~(\ref{eq:partial rate}), the values of ${\cal B}(\DstDpi)=(67.7\pm 0.5)\%$,  ${\cal B}(\DKpi)=(3.947\pm0.030)\%$, and the $B^0$ lifetime $\tau_{B^0}=(1.519 \pm 0.004)$ ps are taken from Ref.~\cite{Workman:2022ynf}. The uncertainties from each source are fully correlated across bins of kinematic variables.

\subsection{Dependence of signal model }

Simulated \bdslnu samples are used to derive migration matrices and efficiency corrections. This introduces a residual dependence on the assumed model into the results. We use the form-factor parameters and $3 \,\sigma$ uncertainties of Ref.~\cite{Ferlewicz:2020lxm} to assess the size of this uncertainty. This systematic uncertainty is smaller than the experimental uncertainties and in most bins does not exceed 1\%. In the $\cos \theta_\ell$ bin of $[-1.0,-0.4]$ it is 4\% and comparable to other uncertainties due to the low reconstruction efficiency. 

%%%%%%%%%%%%%%%%%%%%%%%%%%%%%%%%%%%

\section{Results}\label{sec:Vcb}

By summing the partial decay rates of all kinematic variables we obtain the total rate. The total decay rates averaged over \kinematic are converted to branching fractions using the $B^0$ lifetime. We find
\begin{align}
    {\cal B}(\bdsenu)  &= (4.917\pm 0.032\pm 0.216)\%\, ,\\
    {\cal B}(\bdsmunu) &= (4.926\pm 0.032\pm 0.231)\%\, , 
\end{align}
where the first and second uncertainties are statistical and systematic, respectively. The average is calculated as
\begin{align}
    \BR\, ,
\end{align}
 which is compatible with the current world average: $(4.97\pm 0.12)\%$~\cite{Workman:2022ynf}.

As we investigate partial decay rates of the same dataset, the total decay rate is identical on four projections. These are redundant degrees of freedom in the measured partial decay rates within electrons and muons.
They are removed before analyzing the observed distributions: we calculate normalized partial decay rates $\Delta\Gamma/\Gamma$ and exclude the last bin of each kinematic variable in the determination of form factors and \Vcb. 

We average the electron and muon rates and analyze the observed normalized decay rate $\Delta\Gamma^{\text{obs}}_i/\Gamma^{\text{obs}}$ and total rate $\Gamma^{\text{obs}}$ by constructing a $\chi^2$ function of the form

\begin{linenomath*}
\begin{align}\label{eq:chi2 vcb}
    \chi^2 = &\sum_{i,j}^{34}\left( \frac{\Delta\Gamma^{\text{obs}}_i}{\Gamma^{\text{obs}}} -\frac{\Delta\Gamma_i^{\text{pre}}}{\Gamma^{\text{pre}}} \right) C_{ij}^{-1} \left( \frac{\Delta\Gamma_j^{\text{obs}}}{\Gamma^{\text{obs}}} -\frac{\Delta\Gamma^{\text{pre}}_j}{\Gamma^{\text{pre}}} \right) \\ \nonumber
    &\qquad + \frac{(\Gamma^{\text{obs}}-\Gamma^{\text{pre}})^2}{\sigma_\Gamma^2},
\end{align}
\end{linenomath*}
where $i$ and $j$ denote the indices of the bins in the observables \kinematic, and $\Delta\Gamma_i^{\text{pre}}/\Gamma^{\text{pre}}$ and $\Gamma^{\text{pre}}$ are the predicted values expressed as functions of the form-factor parameters and \Vcb~\cite{Caprini:1997mu,Boyd:1997kz,Boyd:1995sq}.  Further, $C$ is the covariance matrix on the normalized rates, and $\sigma_\Gamma$ is the uncertainty on the total rate.

The input parameters used in the measurement, e.g., $G_F$, $B$-meson mass, and others are summarized in Appendix \ref{app:parameters}. The expansion of BGL form factors must be truncated at a given order. For this we use a nested hypothesis test as proposed in Ref.~\cite{Bernlochner:2019ldg}. We accept a more complex model with an additional expansion parameter over a simpler one if the improvement in $\chi^2$ is one unit or greater. 
We conduct additional testing to ensure that the incorporation of the new expansion parameter does not result in correlations exceeding 95\% among any of the fitted parameters. This precaution is taken to prevent overfitting and the emergence of blind directions, i.e., regions within the parameter space where the $\chi^2$ is approximately uniform.
We identify $n_a = 1$, $n_b = 2$, $n_c = 2$ and in the fits absorb \Vcb  into the fitted expansion coefficients $x_i=\{a_i, b_i, c_i\}$,
\begin{align}\label{eq:tilde par}
\tilde x_i = \Vcb\, \eta_{\mathrm{EW}} \,  x_i  \, .
\end{align}
The obtained values and correlations are listed in Table~\ref{tab:fit BGL} and \Vcb is determined with the relationship
\begin{align}\label{eq:Vcb b0}
    |V_{cb}|\eta_{\mathrm{EW}}{\cal F}(1)=\frac{1}{\sqrt{m_Bm_{D^*}}}\left(\frac{|\tilde{b}_0|}{P_f(0)\phi_f(0)}\right) \, .
\end{align}
Using ${\cal F}(1)=0.906\pm 0.013$~\cite{FermilabLattice:2014ysv} and $\eta_{\mathrm{EW}} = 1.0066$~\cite{Sirlin:1981ie} we determine
\begin{align}
    \VcbBGL.
\end{align}
where the first, second, and third contributions to the uncertainty are statistical, systematic, and from the prediction of ${\cal F}(1)$, respectively. We find a $p$ value of $15$\% for the fit.

\begin{figure*}
\centering
\subfigure{\includegraphics[width=0.45\textwidth, trim = 0.42cm 0.10cm 1.0cm 0.45cm, clip]{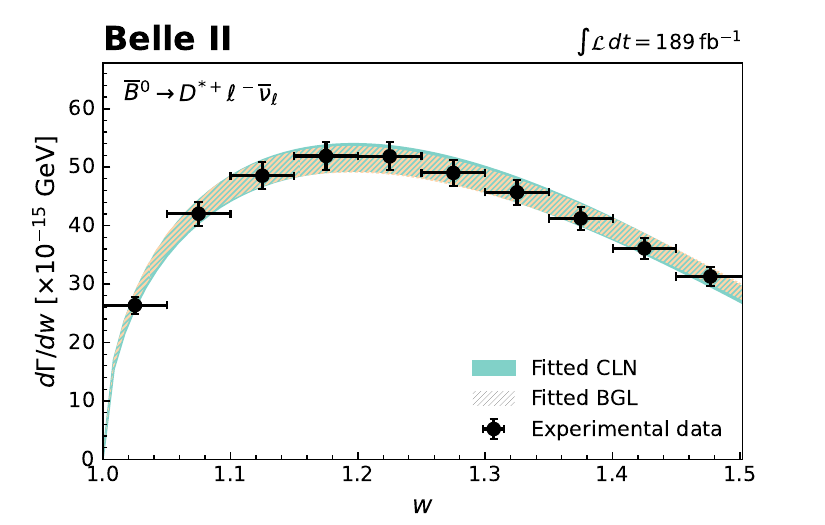}} 
\subfigure{\includegraphics[width=0.45\textwidth, trim = 0.42cm 0.10cm 1.0cm 0.45cm, clip]{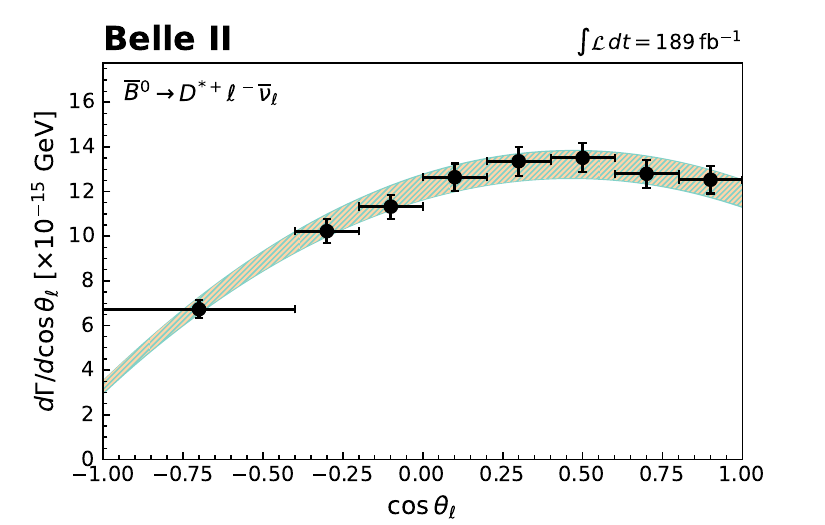}}\\
\subfigure{\includegraphics[width=0.45\textwidth, trim = 0.42cm 0.10cm 1.0cm 0.45cm, clip]{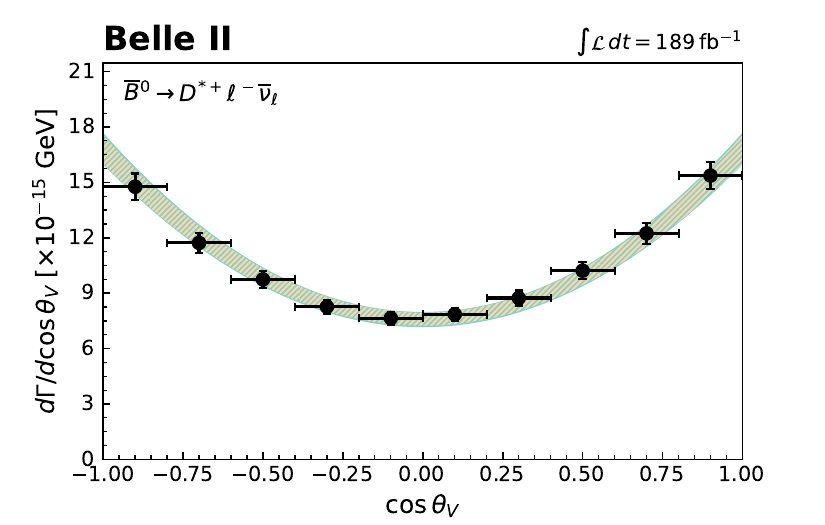}}
\subfigure{\includegraphics[width=0.44\textwidth, trim = 0.42cm 0.10cm 1.0cm 0.45cm, clip]{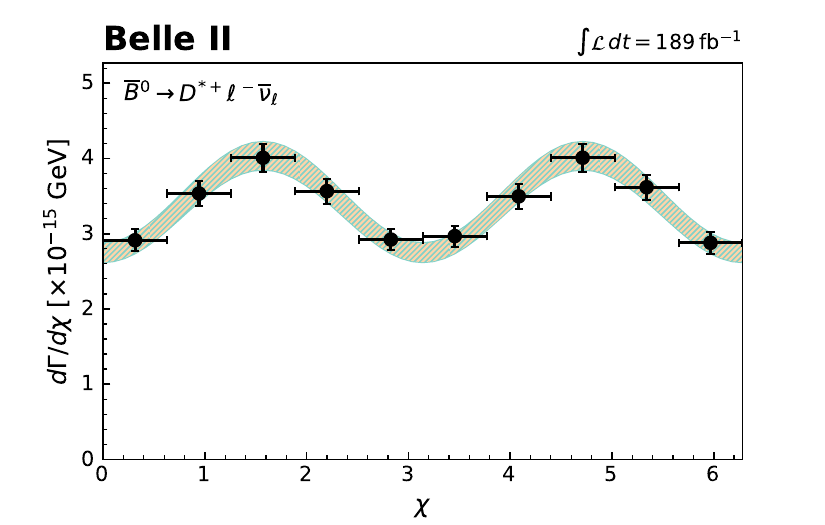}}
\caption{Comparison of the fitted partial decay rates with $1\sigma$ uncertainties in the BGL and CLN parametrizations to the unfolded experimental data (shown as points with error bars). Note that the BGL (hatched) band almost completely overlaps the CLN (solid) band.}
\label{fig:fitted BGL & CLN}
\end{figure*}

Fitting the normalized decay rates and the total decay rate with the CLN parametrization we find
\begin{align}
    \VcbCLN \, ,
\end{align}
with a $p$ value of $16$\%. The fitted parameters and correlations are listed in Table~\ref{tab:fit CLN}. Figure~\ref{fig:fitted BGL & CLN} compares the measured partial decay rates with the fitted differential decay rates. These spectra have very similar shapes for both parametrizations. 

\begin{table}[t]
\begin{center}
\caption{Results of the determination of the BGL expansion coefficients and their correlations.
}\label{tab:fit BGL}
\begin{tabular}{cS[table-format=-1.3(1)]S[table-format=-1.2]S[table-format=-1.2]S[table-format=-1.2]S[table-format=-1.2]c}
\hline\hline
{} &          {Value} & \multicolumn{4}{c}{Correlation} &            $\chi^2$/ndf \\
\hline
$\tilde{a}_0\times 10^3$ &   0.88\pm0.05 &            1.00 &  0.26 & -0.28 &  0.19 &  \multirow{4}{*}{39/31} \\
$\tilde{b}_0\times 10^3$ &   0.54\pm0.01 &         0.26 &     1.00 & -0.37 & -0.43 &                         \\
$\tilde{b}_1\times 10^3$ &  -0.31\pm0.30 &        -0.28 & -0.37 &     1.00 &  0.57 &                         \\
$\tilde{c}_1\times 10^3$ &  -0.04\pm0.03 &         0.19 & -0.43 &  0.57 &     1.00 &                         \\
\hline\hline
\end{tabular}
\end{center}
\end{table}

\begin{table}[t]
\begin{center}
\caption{Results of the determination of the CLN parameters, $|V_{cb}|$ and their correlations.
}\label{tab:fit CLN}
\sisetup{uncertainty-mode=separate}
\renewcommand\arraystretch{1.3} 
\begin{tabular}{cS[table-format=2.3(1)]S[table-format=-1.2]S[table-format=-1.2]S[table-format=-1.2]S[table-format=-1.2]c}
\hline\hline
{} &      {Value} & \multicolumn{4}{c}{Correlation} &            $\chi^2$/ndf \\
\hline
$\rho^2$              &   1.22\pm0.05 &            1.00 &  0.36 & -0.81 &  0.29 &  \multirow{4}{*}{39/31} \\
$R_1(1)$              &   1.14\pm0.07 &         0.36 &     1.00 &  -0.60 &  -0.10 &                         \\
$R_2(1)$              &   0.89\pm0.03 &        -0.81 &  -0.60 &     1.00 & -0.08 &                         \\
$|V_{cb}|\times 10^3$ &  40.1\pm1.1 &         0.29 &  -0.10 & -0.08 &     1.00 &                         \\
\hline\hline
\end{tabular}
\end{center}
\end{table}

A breakdown of the systematic uncertainties for both fits is provided in Tables~\ref{tab:syst:combined BGL} and \ref{tab:syst:combined CLN} for the BGL and CLN parametrizations respectively. The largest uncertainty on \Vcb originates from the knowledge of the slow-pion reconstruction efficiency followed by the uncertainty on the external input $f_{+0}$. 

\begin{table}[htbp]
\begin{center}
\caption{Fractional contributions to the uncertainties of the BGL form factors from a fit of the \bdslnu decay. Because of the absorption of \Vcb into the coefficients (see Eq.~(\ref{eq:tilde par})), the fitted parameters $\tilde x_i$ are affected by the uncertainties that only have an impact on the overall normalization.}
\label{tab:syst:combined BGL}
\begin{tabular}{lccS[table-format=2.1]S[table-format=2.1]}
\hline\hline
{} &  ~~$\tilde{a}_0$~~ &  ~~$\tilde{b}_0$~~ &  {~~$\tilde{b}_1$~~} &  {~~$\tilde{c}_1$~~} \\
\hline
Statistical                            &                3.7 &                0.8 &                 65.1 &                 50.8 \\
Background subtraction                 &                2.1 &                0.4 &                 31.3 &                 21.8 \\
Size of simulated samples                      &                1.5 &                0.3 &                 26.4 &                 20.5 \\
Lepton ID efficiency                   &                1.6 &                0.3 &                  3.4 &                  2.8 \\
Tracking of $K$, $\pi$, $\ell$         &                0.4 &                0.4 &                  0.5 &                  0.4 \\
Slow-pion efficiency                   &                1.6 &                1.5 &                 23.8 &                 24.7 \\
$N_{B\Bbar}$                           &                0.8 &                0.8 &                  0.8 &                  0.8 \\
$f_{+0}$                               &                1.3 &                1.3 &                  1.3 &                  1.2 \\
${\cal B}(D^{*+}\rightarrow D^0\pi^+)$ &                0.4 &                0.4 &                  0.4 &                  0.4 \\
${\cal B}(D^{0}\rightarrow K^-\pi^+)$  &                0.4 &                0.4 &                  0.4 &                  0.4 \\
$B^0$ lifetime                         &                0.1 &                0.1 &                  0.1 &                  0.1 \\
Signal modeling                       &                2.3 &                0.5 &                 52.1 &                 35.0 \\ \hline
Total                                  &                5.8 &                2.5 &                 96.0 &                 73.0 \\
\hline\hline
\end{tabular}
\end{center}
\end{table}

\begin{table}[htbp]
\begin{center}
\caption{Fractional contributions to the uncertainties of the CLN form factors from a fit to the \bdslnu decay. The uncertainties originating from tracking efficiency, the number of $B^0$ mesons, the $B^0$ lifetime, and the charm branching fractions only affect the overall normalization but do not contribute to the parameters related to the shape.
}\label{tab:syst:combined CLN}
\begin{tabular}{lccccc}
\hline\hline
{} &  ~$\rho^2$~ &  ~$R_1(1)$~ &  ~$R_2(1)$~ &  ~$|V_{cb}|$~ \\
\hline
Statistical                            &             3.0 &             4.1 &             2.8 &               0.7 \\
Background subtraction                 &             1.4 &             2.2 &             1.2 &               0.3 \\
Size of simulated samples     &             1.2 &             1.7 &             1.1 &               0.3 \\
Lepton ID efficiency                   &             0.2 &             1.6 &             0.1 &               0.3 \\
Slow pion efficiency                   &             1.0 &             0.9 &             0.8 &               1.5 \\
Tracking of $K$, $\pi$, $\ell$         &                 &                 &                 &               0.4 \\
$N_{B\Bbar}$                           &                 &                 &                 &               0.8 \\
$f_{+0}$                               &                 &                 &                 &               1.3 \\
${\cal B}(D^{*+}\rightarrow D^0\pi^+)$ &                 &                 &                 &               0.4 \\
${\cal B}(D^{0}\rightarrow K^-\pi^+)$  &                 &                 &                 &               0.4 \\
$B^0$ lifetime                         &                 &                 &                 &               0.1 \\
Signal modeling                       &             2.6 &             2.6 &             2.0 &               0.5 \\ \hline
Total                                  &             4.5 &             5.9 &             3.9 &               2.4 \\
\hline\hline
\end{tabular}
\end{center}
\end{table}

\subsection{Sensitivity to FNAL/MILC lattice results at nonzero recoil}\label{sec:lattice}

In Ref.~\cite{FermilabLattice:2021cdg}, the Fermilab lattice and MILC (FNAL/MILC) collaborations published predictions for the \bdslnu form factors at nonzero recoil. We compare our data with these predictions using two scenarios:
\begin{itemize}
    \item Inclusion of predictions beyond zero recoil for $h_{A_1}(w)$ at \mbox{$w=[1.03, 1.10, 1.17]$}. This scenario allows a comparison with the zero-recoil result when information on the $w$ dependence of $h_{A_1}$ is included. 
    \item Inclusion of predictions beyond zero recoil for $h_{A_1}(w)$, $R_1(w)$, and $R_2(w)$ at \mbox{$w=[1.03, 1.10, 1.17]$}. This scenario includes the full LQCD information. 
\end{itemize}

To include beyond-zero-recoil information, we add to Eq.~(\ref{eq:chi2 vcb}) a term of the form
\begin{linenomath*}
\begin{equation}
   \chi^2_{\mathrm{LQCD}} = \sum_{ij}(F_i^{\text{LQCD}}-F_i^{\text{pre}})C^{-1}_{ij}(F_j^{\text{LQCD}}-F_j^{\text{pre}})  \, .
\end{equation}
\end{linenomath*}
Here, $F_i^{\text{LQCD}}$ denotes the lattice data on $h_{A_1}(w)$ or on $h_{A_1}(w)$, $R_1(w)$, $R_2(w)$. The parameter $F_i^{\text{pre}}$ represents the corresponding value expressed in terms of form-factor parameters. As we now explicitly include normalization information on the form factors in the fit, we directly fit for the BGL coefficients without absorbing \Vcb and $\eta_{\mathrm{EW}}$. 

\begin{figure}[h!]
    \centering
    \subfigure{\includegraphics[width=0.41\textwidth, trim = 0.46cm 0.00cm 1.0cm 0.45cm, clip]{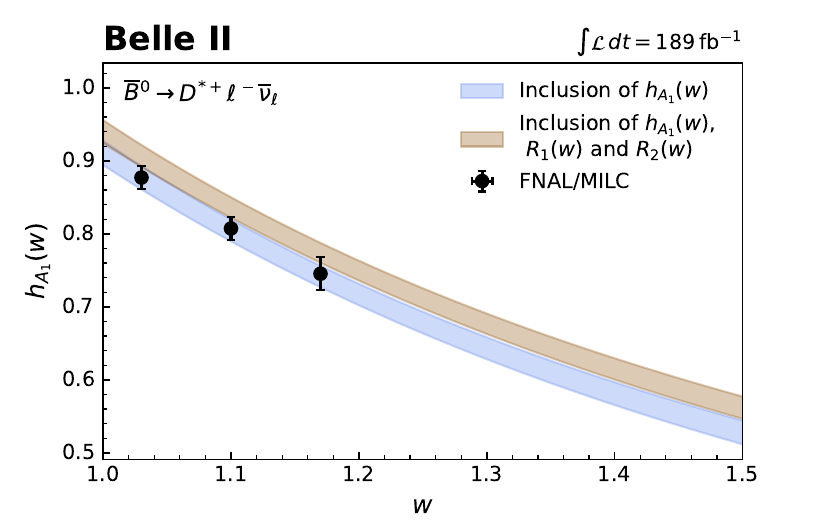}} \\
    \subfigure{\includegraphics[width=0.41\textwidth, trim = 0.46cm 0.00cm 1.0cm 0.45cm, clip]{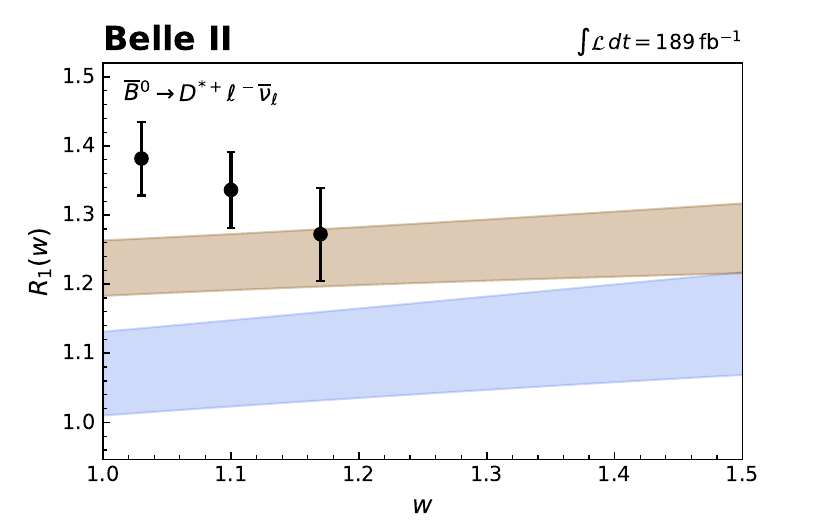}}  \\
    \subfigure{\includegraphics[width=0.41\textwidth, trim = 0.46cm 0.00cm 1.0cm 0.45cm, clip]{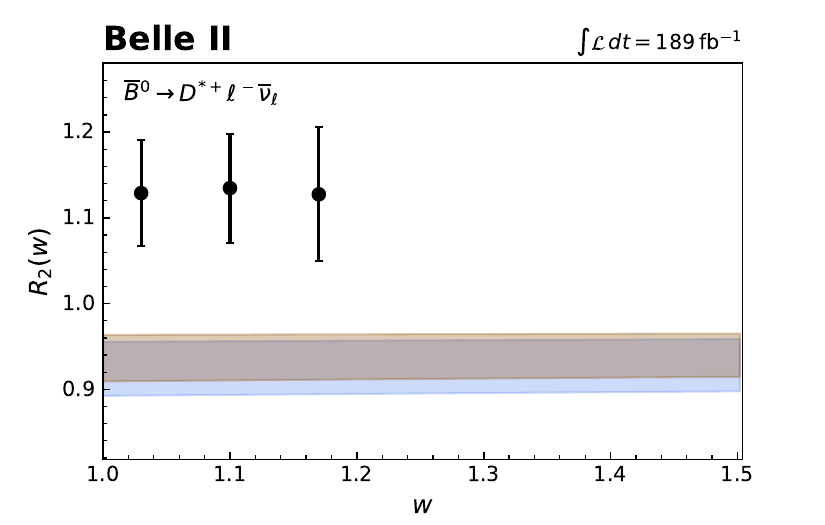}}
    \caption{Comparison of the fitted $h_{A_1}(w)$, $R_1(w)$, and $R_2(w)$ for the BGL fits.}
    \label{fig:BGL_compare_lattice}
\end{figure}

The fitted results in BGL and CLN parametrizations are summarized in Tables~\ref{tab:fit lattice BGL} and \ref{tab:fit lattice CLN}, respectively. The inclusion of beyond-zero-recoil information for $h_{A_1}$ results in a small decrease of the central value for \Vcb if we use the BGL form-factor expansion. The CLN fits show a small increase. The inclusion of the full beyond-zero-recoil information shifts \Vcb significantly and the resulting fit shapes in $h_{A_1}$, $R_1$, and $R_2$ disagree with the FNAL/MILC lattice predictions with a poor $p$ value of $0.04\%$. This is consistent with the results of Ref.~\cite{Belle:2023bwv}. The BGL fits of both scenarios are shown in Fig.~\ref{fig:BGL_compare_lattice} with the nonzero recoil FNAL/MILC predictions of Ref.~\cite{FermilabLattice:2021cdg}. The agreement can be improved if more BGL expansion parameters are included: in Appendix~\ref{app:NHT} we repeat the nested hypothesis test to determine the appropriate truncation order when full lattice information is included and find $n_a = 3$, $n_b = 1$, $n_c = 3$. With 6 expansion coefficients we find a $p$ value of 16\% because of better agreement of $R_1(w)$ and $R_2(w)$ at \mbox{$w=[1.03, 1.10, 1.17]$}.

\begin{table}[htbp]
\begin{center}
\renewcommand\arraystretch{1.3} 
\caption{Values of BGL form factors and \Vcb resulting from a fit that includes nonzero recoil lattice information.
}\label{tab:fit lattice BGL}
\sisetup{uncertainty-mode=separate}
\begin{tabular}{lS[table-format=-2.3(1)]S[table-format=-2.3(1)]}
\hline\hline
{} & \multicolumn{1}{c}{\makecell{Constraints on \\ $h_{A_1}(w)$}} & \multicolumn{1}{c}{\makecell{Constraints on \\ $h_{A_1}(w)$, $R_1(w)$, $R_2(w)$}} \\
\hline
$a_0\times 10^3$      &                                  21.7\pm1.3 &                                         25.6\pm0.8 \\
$b_0\times 10^3$      &                                13.19\pm0.24 &                                       13.61\pm0.23 \\
$b_1\times 10^3$      &                                      -6\pm6 &                                              2\pm6 \\
$c_1\times 10^3$      &                                  -0.9\pm0.7 &                                         -0.0\pm0.7 \\
$|V_{cb}|\times 10^3$ &                                  40.3\pm1.2 &                                         38.3\pm1.1 \\ \hline
$\chi^2$/ndf          &                                     {39/33} &                                            {75/39} \\
$p$ value             &                                      {21\%} &                                              {0.04\%} \\
\hline\hline
\end{tabular}
\end{center}
\end{table}

\begin{table}[htbp]
\begin{center}
\sisetup{uncertainty-mode=separate}
\renewcommand\arraystretch{1.3} 
\caption{Values of CLN form factors and \Vcb resulting from a fit that includes nonzero recoil lattice information. 
}\label{tab:fit lattice CLN}
\begin{tabular}{cS[table-format=2.3(1)]S[table-format=2.3(1)]}
\hline\hline
{} & \multicolumn{1}{c}{\makecell{Constraints on \\ $h_{A_1}(w)$}} & \multicolumn{1}{c}{\makecell{Constraints on \\ $h_{A_1}(w)$, $R_1(w)$, $R_2(w)$}} \\
\hline
$h_{A_1}(1)$          &                                 0.91\pm0.02 &                                        0.94\pm0.02 \\
$\rho^2$              &                                 1.22\pm0.05 &                                        1.21\pm0.04 \\
$R_1(1)$              &                                 1.14\pm0.07 &                                        1.26\pm0.04 \\
$R_2(1)$              &                                 0.88\pm0.03 &                                        0.88\pm0.03 \\
$|V_{cb}|\times 10^3$ &                                  40.3\pm1.2 &                                         38.7\pm1.1 \\ \hline
$\chi^2$/ndf          &                                     {39/33} &                                            {70/39} \\
$p$ value             &                                      {23\%} &                                            {0.2\%} \\
\hline\hline
\end{tabular}
\end{center}
\end{table}

\subsection{Lepton-flavor universality test}\label{sec:A_FB}

We report a value for the ratio of the \bdsenu and \bdsmunu branching fractions
\begin{linenomath*}
\begin{equation}
    R_{e/\mu} = \Remu,
\end{equation}
\end{linenomath*}
where the first contribution to the uncertainty is statistical and the second systematic. The ratio is compatible with the predictions of Refs.~\cite{Bernlochner:2022ywh,Bobeth:2021lya} (see Table~\ref{tab:Summary of LFU} for a summary) assuming LFU and with previous measurements~\cite{Belle:2018ezy,Belle-II:2023qyd}. The fully correlated systematic uncertainties, e.g., the tracking efficiency, the number of $B^0$ mesons, and the branching fractions of the $D^{*+}$ and $D^0$ decays cancel in the ratio.

From the observed partial decay rates on the $\cos\theta_\ell$ projection, we determine the angular asymmetry ${\cal A}_{\mathrm{FB}}$ in the full phase space of $w$, 
\begin{align}
    {\cal A}_{\mathrm{FB}}=\frac{\int^1_0\text{d}\cos\theta_\ell\text{d}\Gamma/\text{d}\cos\theta_\ell-\int^0_{-1}\text{d}\cos\theta_\ell\text{d}\Gamma/\text{d}\cos\theta_\ell}{\int^1_0\text{d}\cos\theta_\ell\text{d}\Gamma/\text{d}\cos\theta_\ell+\int^0_{-1}\text{d}\cos\theta_\ell\text{d}\Gamma/\text{d}\cos\theta_\ell}.
\end{align}
With ${\cal A}_{\mathrm{FB}}$ we test LFU using the difference
\begin{align}
    \Delta{\cal A}_{\mathrm{FB}}={\cal A}_{\mathrm{FB}}^\mu-{\cal A}_{\mathrm{FB}}^e \, .
\end{align}
We find
\begin{align}
    {\cal A}_{\mathrm{FB}}^e &= 0.228 \pm 0.012 \pm 0.018\, , \\
    {\cal A}_{\mathrm{FB}}^\mu &= 0.211 \pm 0.011 \pm 0.021\, ,
\end{align}
and 
\begin{equation}
        \DAFB\, .
\end{equation}
The correlated uncertainties between the \bdsenu and \bdsmunu decays, e.g., the number of $B^0$ mesons, the $B^0$ lifetime, and others cancel in $\Delta{\cal A}_{\mathrm{FB}}$. Note that due to the selection requirement on the lepton momentum in the c.m.\ system, the unfolded yields in the negative $\cos\theta_\ell$ region require a large correction based on the SM assumption. Consequently, the measured value of $\Delta{\cal A}_{\mathrm{FB}}$ would change in the presence of non-SM physics, and should only be used to check for consistency with the SM expectation.

To minimize the extrapolation, we also measure ${\cal A}_{\mathrm{FB}}$ in the phase space of $p_\ell^{B} > \SI{1.2}{\GeV/}c$, with $p_\ell^{B}$ denoting the lepton momentum in the $B$ meson rest frame. We find
\begin{align}
    {\cal A}_{\mathrm{FB}}^e(p_\ell^{B} > \SI{1.2}{\GeV/}c) &= 0.611 \pm 0.006 \pm 0.005\, , \\
    {\cal A}_{\mathrm{FB}}^\mu(p_\ell^{B} > \SI{1.2}{\GeV/}c) &= 0.604 \pm 0.006 \pm 0.008\, , \\
    \Delta {\cal A}_{\mathrm{FB}} (p_\ell^{B} > \SI{1.2}{\GeV/}c) &=  \left(-7 \pm 9 \pm 9 \right) \times 10^{-3} \, .
\end{align}

From the observed $\cos\theta_V$ distribution, we determine the longitudinal $D^*$ polarization fraction $F_L$ via
\begin{align}\label{eq:F_L}
    \frac{1}{\Gamma}\frac{\text{d}\Gamma}{\text{d}\cos\theta_V}=\frac{3}{2}\left(F_L\cos^2\theta_V+\frac{1-F_L}{2}\sin^2\theta_V\right) \, ,
\end{align}
and we find
\begin{linenomath*}
\begin{align}
    F_L^{e} &= 0.520 \pm 0.005 \pm 0.005\, ,\\
    F_L^{\mu} &= 0.527 \pm 0.005 \pm 0.005\, , 
\end{align}
\end{linenomath*}
and
\begin{equation}
    \DFL  \,  ,
\end{equation}
with $\Delta F_L = F_L^{\mu} - F_L^{e}$. The correlated uncertainties between the \bdsenu and \bdsmunu decays cancel in $\Delta F_L$.

The resulting angular asymmetry and longitudinal polarization for \bdsenu and \bdsmunu decays and their difference between the $e$ channel and $\mu$ channel agree with the SM predictions of Refs.~\cite{Bobeth:2021lya,Bernlochner:2022ywh}, which are summarized in Table~\ref{tab:Summary of LFU}. Note that ${\cal A}_{\mathrm{FB}}$ in Ref.~\cite{Bobeth:2021lya} is determined from a slightly reduced phase space corresponding to $1.0 < w < 1.5$. However, the impact of this restriction on the SM expectations is of order $10^{-4}$~\cite{Bernlochner:2022ywh}. 

Our values are compatible with the determination of $\Delta{\cal A}_{\mathrm{FB}}$ and $\Delta F_L$ of Refs.~\cite{Bobeth:2021lya,Belle:2018ezy} within 2.3 
and 1.2 standard deviations, respectively. Recently Ref.~\cite{Belle:2023bwv} also determined these quantities and we observe good agreement for ${\cal A}_{\mathrm{FB}}$ and $F_L$ for electron and muon final states and their differences. 

\begin{table}[t]
\begin{center}
\caption{Summary of the SM predictions taken from Ref.~\cite{Bernlochner:2022ywh,Bobeth:2021lya} for LFU tests. Note that the $F_L$ in Refs.~\cite{Bernlochner:2022ywh} is only reported with $m_\ell=0$ for the light leptons $\ell=e, \mu$.
}\label{tab:Summary of LFU}
\sisetup{uncertainty-mode=separate}
\renewcommand\arraystretch{1.3} 
\begin{tabular}{cS[table-format=-1.5(1)]S[table-format=-1.5(1)]cccc}
\hline\hline
{} &      {Ref.~\cite{Bernlochner:2022ywh}} & {Ref.~\cite{Bobeth:2021lya}}  \\ \hline
$R_{e/\mu}$ &  1.0041(1) &        1.0026\pm0.0001    \\
${\cal A}^e_{\mathrm{FB}}$              &  0.244(4)  &            0.204\pm0.012  \\
${\cal A}^\mu_{\mathrm{FB}}$              &   
0.239(4)  &         0.198\pm0.012  \\
$\Delta{\cal A}_{\mathrm{FB}} \times 10^{3}$              &   -5.7(1) &         -5.33\pm0.24      \\
$F_L^e$              &  0.516\pm0.003   &            0.541\pm0.011    \\
$F_L^\mu$              &   0.516\pm0.003  &         0.542\pm0.012  \\
$\Delta F_L \times 10^{4}$           &   1.2\pm0.1 &         5.43\pm0.36    \\
\hline\hline
\end{tabular}
\end{center}
\end{table}

%%%%%%%%%%%%%%%%%%%%%%%%%%%%%%%%%%%

\section{Summary and conclusion}\label{sec:summary}

In this paper we present a measurement of partial decay rates of \bdsenu and \bdsmunu channels using a sample corresponding to \lumi of Belle~II data. We unfold the measured partial decay rates to account for detector and efficiency effects and analyze them to determine form factors and the value of \Vcb. Using the BGL parametrization we find 
\begin{align}
    \VcbBGL\, ,
\end{align}
which is in good agreement with the world average of the exclusive approach and the inclusive determination of Refs.~\cite{Bordone:2021oof,Bernlochner:2022ucr}. Using the CLN parametrization results in a similar, but lower value,
\begin{align}
    \VcbCLN\, .
\end{align}
The obtained \Vcb values of BGL and CLN parametrizations agree with the recent Belle measurement~\cite{Belle:2023bwv}. 
The slope difference of the form factor near zero recoil is the reason for the small upward shift of the BGL-based \Vcb\ value in comparison to the CLN result. The slopes of $\mathcal{F}(w)$ at zero recoil are found to be $\mathcal{F}'|_{w=1} = -2.03 \pm 0.11$ (BGL) and $-1.86 \pm 0.08$ (CLN). Both measured values of \Vcb are compatible with the exclusive and inclusive world averages~\cite{HFLAV:2022pwe,Bordone:2021oof} within 1.5 or 1.3 (BGL) and 1.1 or 1.6 (CLN) standard deviations. The precision of the results is limited by the knowledge of the slow-pion efficiency, which can be improved with larger data sets. The current world average of \Vcb\ is dominated by measurements of $\bdslnu$ decays within the CLN parametrization. Using the values of the CLN parameters taken from Ref.~\cite{HFLAV:2022pwe}, one finds $\mathcal{F}'|_{w=1} -1.54 \pm 0.05$, which is $\simeq 3$ standard deviations higher than the CLN-based slope reported in this paper.

We also test the impact of including FNAL/MILC lattice predictions at nonzero recoil from Ref.~\cite{FermilabLattice:2021cdg} with the same order of BGL expansion in two scenarios: when nonzero recoil information for $h_{A_1}$ is included, the resulting value of \Vcb decreases slightly. With the full information on all form factors included, the resulting functional dependence on $h_{A_1}$, $R_1$ and $R_2$ is in tension with the FNAL/MILC lattice predictions, and the BGL fit results in a poor $p$ value of 0.04\% if one uses the same number of BGL expansion parameters as for the data only fit. Repeating the fits with more parameters can provide better agreement, but the predicted functional dependence of $R_2(w)$ is in tension with the FNAL/MILC LQCD predictions. 

We test the electron-muon LFU by determining the ratio of branching fractions. The result
\begin{equation}
    R_{e/\mu} = \Remu ,
\end{equation}
is in good agreement with unity. To further test LFU, we also measure the forward-backward asymmetry and the $D^{*+}$ polarization, and find \DAFB and \DFL in good agreement with the SM expectations. 

%%%%%%%%%%%%%%%%%%%%%%%%%%%%%%%%%%%

\acknowledgments
We thank S. Cal\`{o} for his contributions to the reconstruction methods for the kinematic variables.  C. Lyu is dedicating this paper to his grandmother X. Wang, who sadly passed away during the preparation of this manuscript. % Policy from October 20, 2022
This work, based on data collected using the Belle II detector, which was built and commissioned prior to March 2019, was supported by
%Armenia
Science Committee of the Republic of Armenia Grant No.~20TTCG-1C010;
%Australia
Australian Research Council and Research Grants
No.~DP200101792, % Jackson
No.~DP210101900, % Urquijo
No.~DP210102831, % Sevior
No.~DE220100462, % Hsu
No.~LE210100098, % Infrastructure
and
No.~LE230100085; % Infrastructure
%Austria
Austrian Federal Ministry of Education, Science and Research,
Austrian Science Fund
No.~P~31361-N36
and
No.~J4625-N,
and
Horizon 2020 ERC Starting Grant No.~947006 ``InterLeptons'';
%Canada
Natural Sciences and Engineering Research Council of Canada, Compute Canada and CANARIE;
%China
National Key R\&D Program of China under Contract No.~2022YFA1601903,
National Natural Science Foundation of China and Research Grants
No.~11575017,
No.~11761141009,
No.~11705209,
No.~11975076,
No.~12135005,
No.~12150004,
No.~12161141008,
and
No.~12175041,
and Shandong Provincial Natural Science Foundation Project~ZR2022JQ02;
%Czech Republic
the Czech Science Foundation Grant No.~22-18469S;
%EU
European Research Council, Seventh Framework PIEF-GA-2013-622527,
Horizon 2020 ERC-Advanced Grants No.~267104 and No.~884719,
Horizon 2020 ERC-Consolidator Grant No.~819127,
Horizon 2020 Marie Sklodowska-Curie Grant Agreement No.~700525 ``NIOBE''
and
No.~101026516,
and
Horizon 2020 Marie Sklodowska-Curie RISE project JENNIFER2 Grant Agreement No.~822070 (European grants);
%France
L'Institut National de Physique Nucl\'{e}aire et de Physique des Particules (IN2P3) du CNRS
and
L'Agence Nationale de la Recherche (ANR) under grant ANR-21-CE31-0009 (France);
%Germany
BMBF, DFG, HGF, MPG, and AvH Foundation (Germany);
%India
Department of Atomic Energy under Project Identification No.~RTI 4002,
Department of Science and Technology,
and
UPES SEED funding programs
No.~UPES/R\&D-SEED-INFRA/17052023/01 and
No.~UPES/R\&D-SOE/20062022/06 (India);
%Israel
Israel Science Foundation Grant No.~2476/17,
U.S.-Israel Binational Science Foundation Grant No.~2016113, and
Israel Ministry of Science Grant No.~3-16543;
%Italy
Istituto Nazionale di Fisica Nucleare and the Research Grants BELLE2;
%Japan
Japan Society for the Promotion of Science, Grant-in-Aid for Scientific Research Grants
No.~16H03968,
No.~16H03993,
No.~16H06492,
No.~16K05323,
No.~17H01133,
No.~17H05405,
No.~18K03621,
No.~18H03710,
No.~18H05226,
No.~19H00682, % Niigata
No.~22H00144,
No.~22K14056,
No.~22K21347,
No.~23H05433,
No.~26220706,
and
No.~26400255,
the National Institute of Informatics, and Science Information NETwork 5 (SINET5), 
and
the Ministry of Education, Culture, Sports, Science, and Technology (MEXT) of Japan;  
%Korea
National Research Foundation (NRF) of Korea Grants
No.~2016R1\-D1A1B\-02012900,
No.~2018R1\-A2B\-3003643,
No.~2018R1\-A6A1A\-06024970,
No.~2019R1\-I1A3A\-01058933,
No.~2021R1\-A6A1A\-03043957,
No.~2021R1\-F1A\-1060423,
No.~2021R1\-F1A\-1064008,
No.~2022R1\-A2C\-1003993,
and
No.~RS-2022-00197659,
Radiation Science Research Institute,
Foreign Large-Size Research Facility Application Supporting project,
the Global Science Experimental Data Hub Center of the Korea Institute of Science and Technology Information
and
KREONET/GLORIAD;
%Malaysia
Universiti Malaya RU grant, Akademi Sains Malaysia, and Ministry of Education Malaysia;
%Mexico
% CINVESTAV-IPN, UNAM, UAS, BUAP and CONACYT are funded under
Frontiers of Science Program Contracts
No.~FOINS-296,
No.~CB-221329,
No.~CB-236394,
No.~CB-254409,
and
No.~CB-180023, and SEP-CINVESTAV Research Grant No.~237 (Mexico);
%Poland
the Polish Ministry of Science and Higher Education and the National Science Center;
%Russia
the Ministry of Science and Higher Education of the Russian Federation,
Agreement No.~14.W03.31.0026, and
the HSE University Basic Research Program, Moscow;
%Saudi Arabia
University of Tabuk Research Grants
No.~S-0256-1438 and No.~S-0280-1439 (Saudi Arabia);
%Slovenia
Slovenian Research Agency and Research Grants
No.~J1-9124
and
No.~P1-0135;
%Spain
Agencia Estatal de Investigacion, Spain
Grant No.~RYC2020-029875-I
and
Generalitat Valenciana, Spain
Grant No.~CIDEGENT/2018/020;
%Taiwan
National Science and Technology Council,
and
Ministry of Education (Taiwan);
%Thailand
Thailand Center of Excellence in Physics;
%Turkey
TUBITAK ULAKBIM (Turkey);
%Ukraine
National Research Foundation of Ukraine, Project No.~2020.02/0257,
and
Ministry of Education and Science of Ukraine;
%USA
the U.S. National Science Foundation and Research Grants
No.~PHY-1913789 % Indiana CEEM
and
No.~PHY-2111604, % Luther
and the U.S. Department of Energy and Research Awards
No.~DE-AC06-76RLO1830, % PNNL
No.~DE-SC0007983, % Wayne State
No.~DE-SC0009824, % Florida
No.~DE-SC0009973, % VPI
No.~DE-SC0010007, % Duke
No.~DE-SC0010073, % South Carolina
No.~DE-SC0010118, % Carnegie Mellon
No.~DE-SC0010504, % Hawaii
No.~DE-SC0011784, % Cincinnati
No.~DE-SC0012704, % BNL
No.~DE-SC0019230, % Duke
No.~DE-SC0021274, % Mississippi
No.~DE-SC0022350, % Louisville
No.~DE-SC0023470; % South Alabama
%last group
and
%Vietnam
the Vietnam Academy of Science and Technology (VAST) under Grant No.~DL0000.05/21-23.

% Policy from October 20, 2022
These acknowledgements are not to be interpreted as an endorsement of any statement made
by any of our institutes, funding agencies, governments, or their representatives.

We thank the SuperKEKB team for delivering high-luminosity collisions;
the KEK cryogenics group for the efficient operation of the detector solenoid magnet;
the KEK computer group and the NII for on-site computing support and SINET6 network support;
and the raw-data centers at BNL, DESY, GridKa, IN2P3, INFN, and the University of Victoria for off-site computing support.

\bibliographystyle{apsrev4-1}
\bibliography{Dstlnu}

%%%%%%%%%%%%%%%%%%%%%%%%%%%%%%%%%%%

\clearpage

\onecolumngrid
\begin{appendix}

\FloatBarrier
\section{SUMMARY OF FITTED YIELDS}\label{app:yields}

The obtained signal yields in bins of kinematic variables are summarized in Table~\ref{tab:fitted yields}.

\begin{table*}[h!]
\renewcommand\arraystretch{1.2}
\caption{Obtained yields in bins of kinematic variables. The uncertainty is statistical only.}
\label{tab:fitted yields}
\sisetup{uncertainty-mode=separate,table-number-alignment=center}
\begin{tabular}{clS[table-format=4.1(4)]S[table-format=4.1(4)]}
\hline\hline
      \multicolumn{1}{c}{Variable} & \multicolumn{1}{c}{Bin} & \multicolumn{1}{c}{\bdsenu} & \multicolumn{1}{c}{\bdsmunu} \\ \hline
 
              \multirow{10}{*}{$w$} &            $[1.00, 1.05)$ &                   2004\pm84 &                    2055\pm85 \\
                                    &            $[1.05, 1.10)$ &                  4519\pm120 &                   4927\pm122 \\
                                    &            $[1.10, 1.15)$ &                  6252\pm126 &                   6807\pm126 \\
                                    &            $[1.15, 1.20)$ &                  7433\pm127 &                   7806\pm129 \\
                                    &            $[1.20, 1.25)$ &                  7636\pm125 &                   8180\pm128 \\
                                    &            $[1.25, 1.30)$ &                  7326\pm121 &                   7686\pm123 \\
                                    &            $[1.30, 1.35)$ &                  6531\pm112 &                   7016\pm118 \\
                                    &            $[1.35, 1.40)$ &                  5615\pm106 &                   6091\pm110 \\
                                    &            $[1.40, 1.45)$ &                  4699\pm100 &                   4924\pm109 \\
                                    &            $[1.45, 1.51)$ &                   3352\pm99 &                   3626\pm110 \\ \hline
 \multirow{8}{*}{$\cos\theta_\ell$} &          $[-1.00, -0.40)$ &                   1811\pm82 &                    1990\pm87 \\
                                    &          $[-0.40, -0.20)$ &                   2136\pm81 &                    2368\pm85 \\
                                    &           $[-0.20, 0.00)$ &                  5085\pm120 &                   5214\pm127 \\
                                    &            $[0.00, 0.20)$ &                  8901\pm154 &                   9679\pm159 \\
                                    &            $[0.20, 0.40)$ &                 10163\pm145 &                  10056\pm146 \\
                                    &            $[0.40, 0.60)$ &                 10020\pm140 &                  10728\pm141 \\
                                    &            $[0.60, 0.80)$ &                  9286\pm132 &                  10312\pm139 \\
                                    &            $[0.80, 1.00)$ &                  8018\pm121 &                   8872\pm129 \\ \hline
   \multirow{10}{*}{$\cos\theta_V$} &          $[-1.00, -0.80)$ &                  7930\pm126 &                   8802\pm131 \\
                                    &          $[-0.80, -0.60)$ &                  6963\pm122 &                   7217\pm126 \\
                                    &          $[-0.60, -0.40)$ &                  6209\pm117 &                   6481\pm121 \\
                                    &          $[-0.40, -0.20)$ &                  5297\pm113 &                   5762\pm117 \\
                                    &           $[-0.20, 0.00)$ &                  5025\pm112 &                   5321\pm114 \\
                                    &            $[0.00, 0.20)$ &                  4887\pm110 &                   5194\pm112 \\
                                    &            $[0.20, 0.40)$ &                  4720\pm108 &                   5235\pm110 \\
                                    &            $[0.40, 0.60)$ &                  4921\pm106 &                   5234\pm109 \\
                                    &            $[0.60, 0.80)$ &                  4800\pm103 &                   4935\pm107 \\
                                    &            $[0.80, 1.00)$ &                   4650\pm98 &                   5038\pm104 \\ \hline
           \multirow{10}{*}{$\chi$} &            $[0.00, 0.63)$ &                  4660\pm107 &                   4992\pm112 \\
                                    &            $[0.63, 1.26)$ &                  5458\pm115 &                   6033\pm117 \\
                                    &            $[1.26, 1.88)$ &                  6542\pm117 &                   6741\pm119 \\
                                    &            $[1.88, 2.51)$ &                  5885\pm114 &                   6358\pm118 \\
                                    &            $[2.51, 3.14)$ &                  5063\pm107 &                   5362\pm113 \\
                                    &            $[3.14, 3.77)$ &                  5222\pm109 &                   5516\pm112 \\
                                    &            $[3.77, 4.40)$ &                  5815\pm112 &                   6188\pm116 \\
                                    &            $[4.40, 5.03)$ &                  6361\pm117 &                   6930\pm120 \\
                                    &            $[5.03, 5.65)$ &                  5653\pm112 &                   6240\pm116 \\
                                    &            $[5.65, 6.28)$ &                  4726\pm106 &                   4831\pm110 \\
\hline\hline
\end{tabular}
\end{table*}

\clearpage
\FloatBarrier
\section{MIGRATION MATRICES FOR THE MUON CHANNEL}\label{app:migration}

\begin{figure*}[h!]
    \centering
    \subfigure{\includegraphics[width=0.35\textwidth]{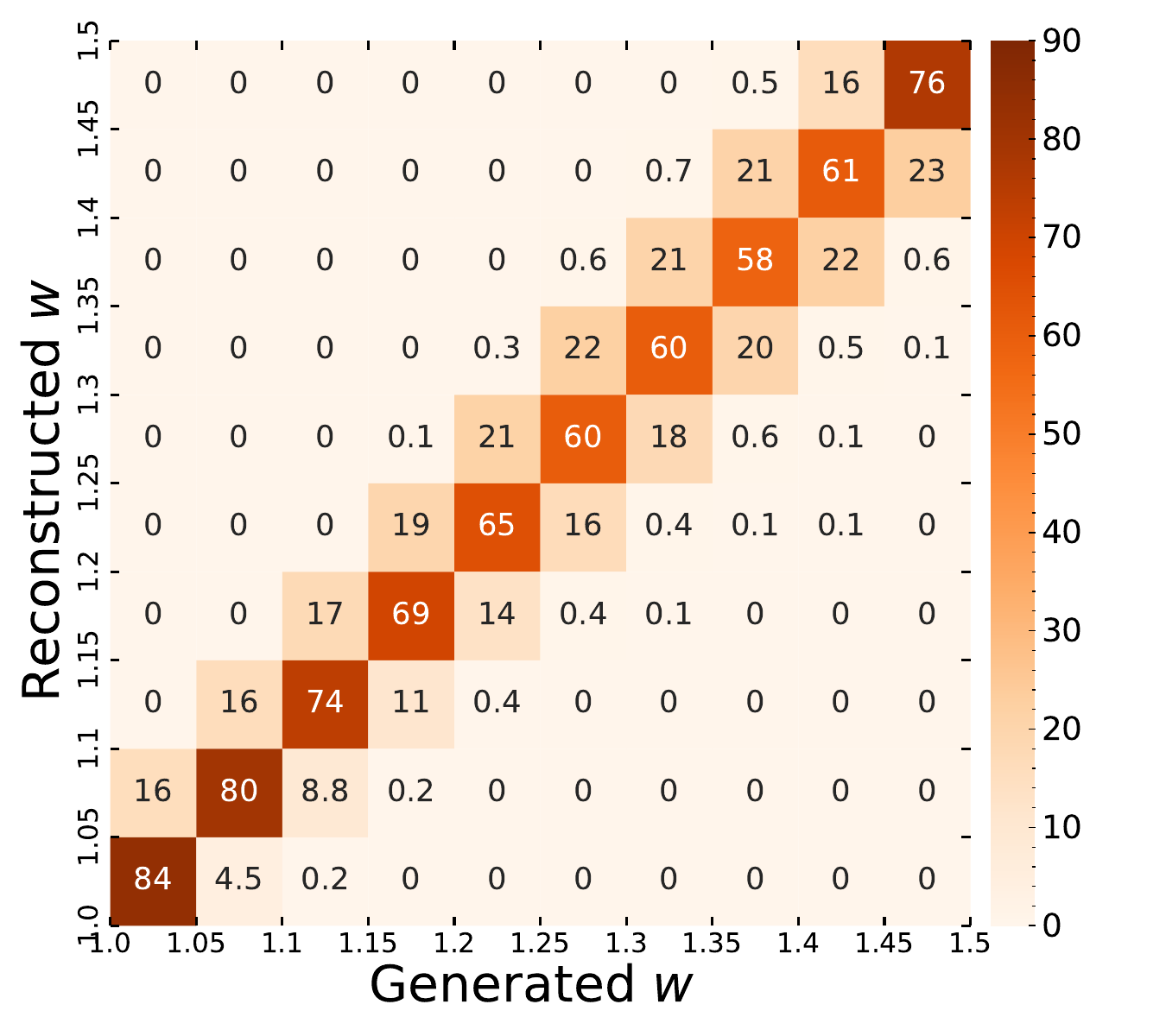}} %, trim = 0.42cm 0.40cm 1.45cm 0.45cm, clip
    \subfigure{\includegraphics[width=0.35\textwidth]{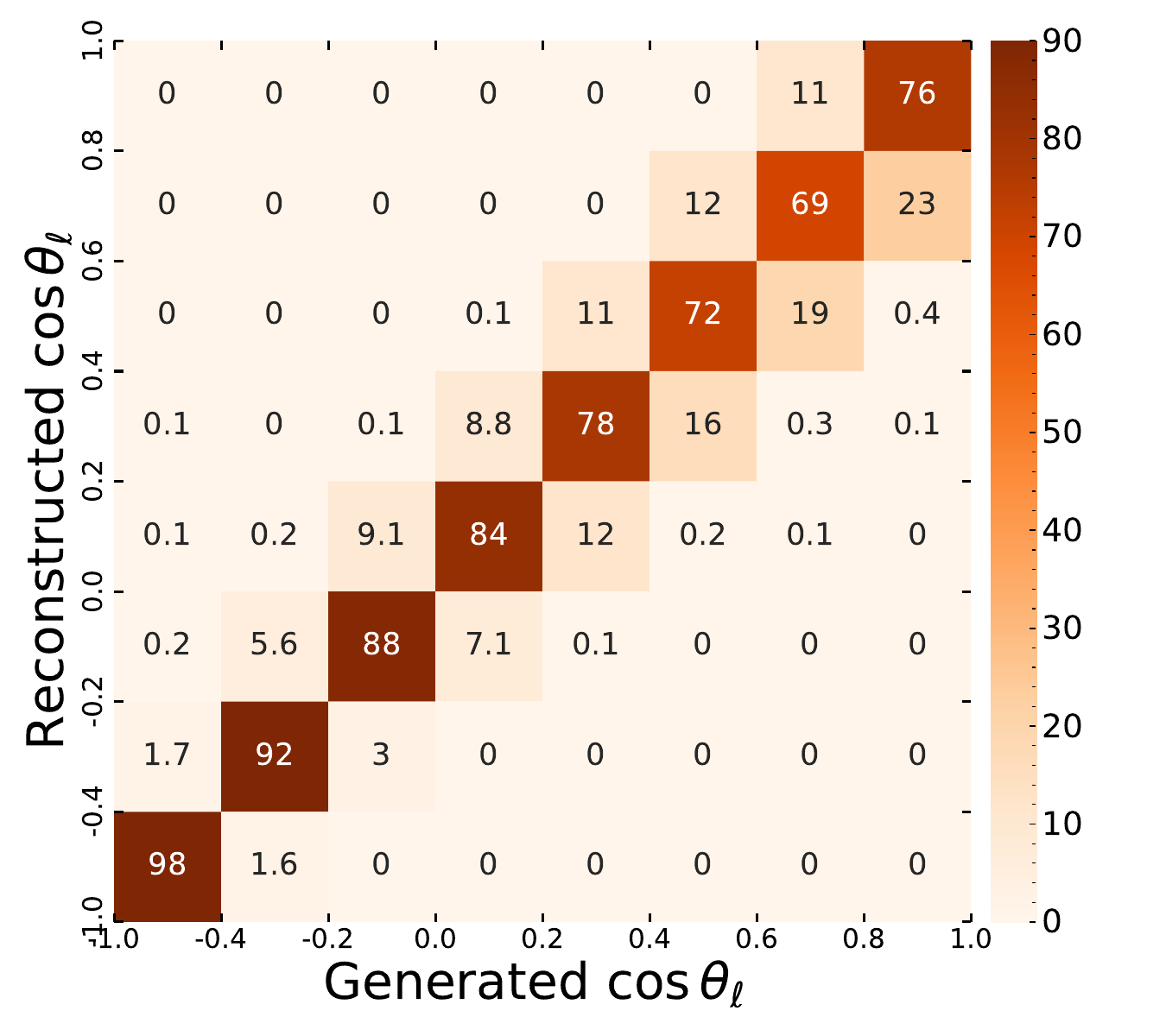}} \\
    \subfigure{\includegraphics[width=0.35\textwidth]{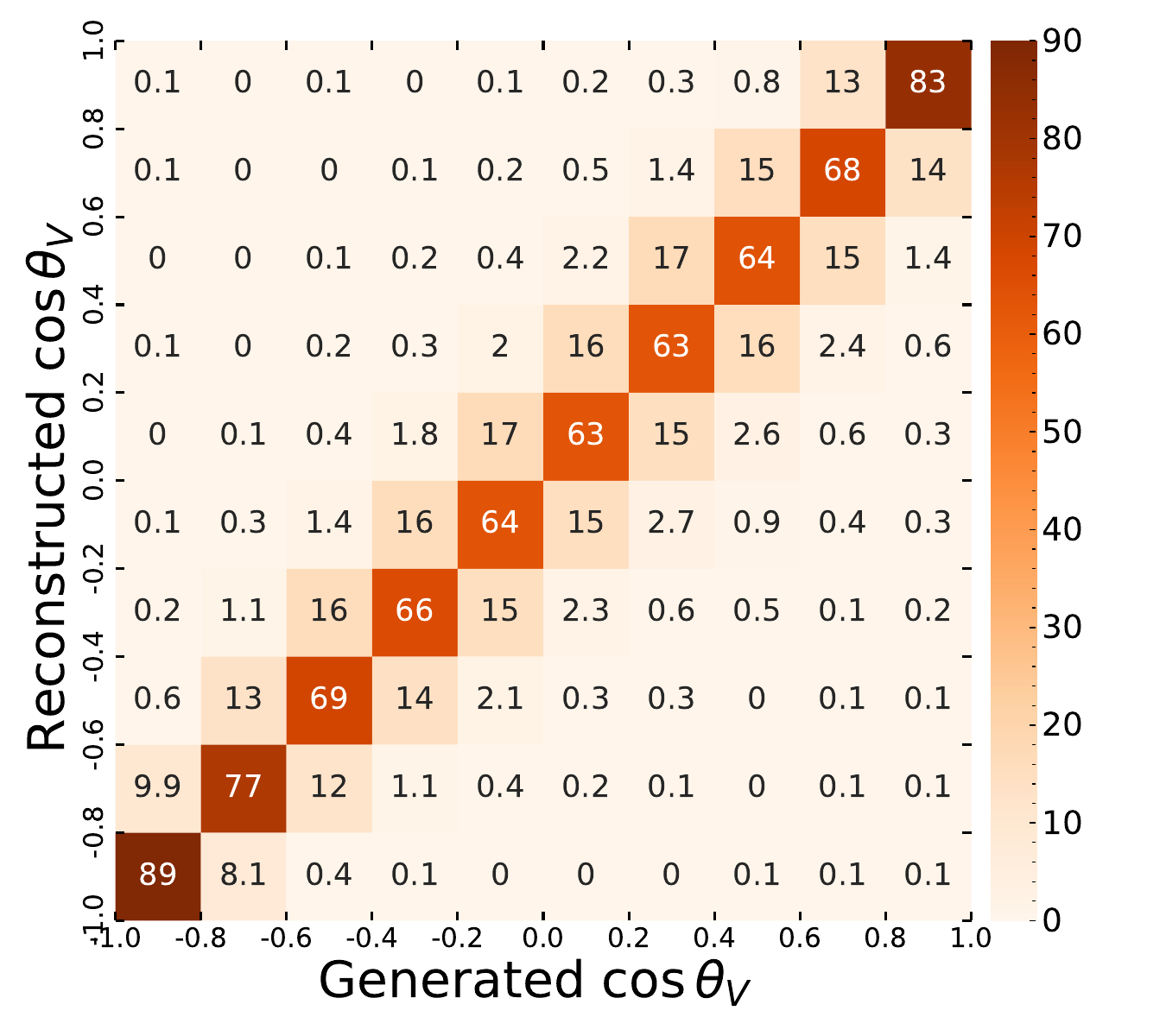}}
    \subfigure{\includegraphics[width=0.35\textwidth]{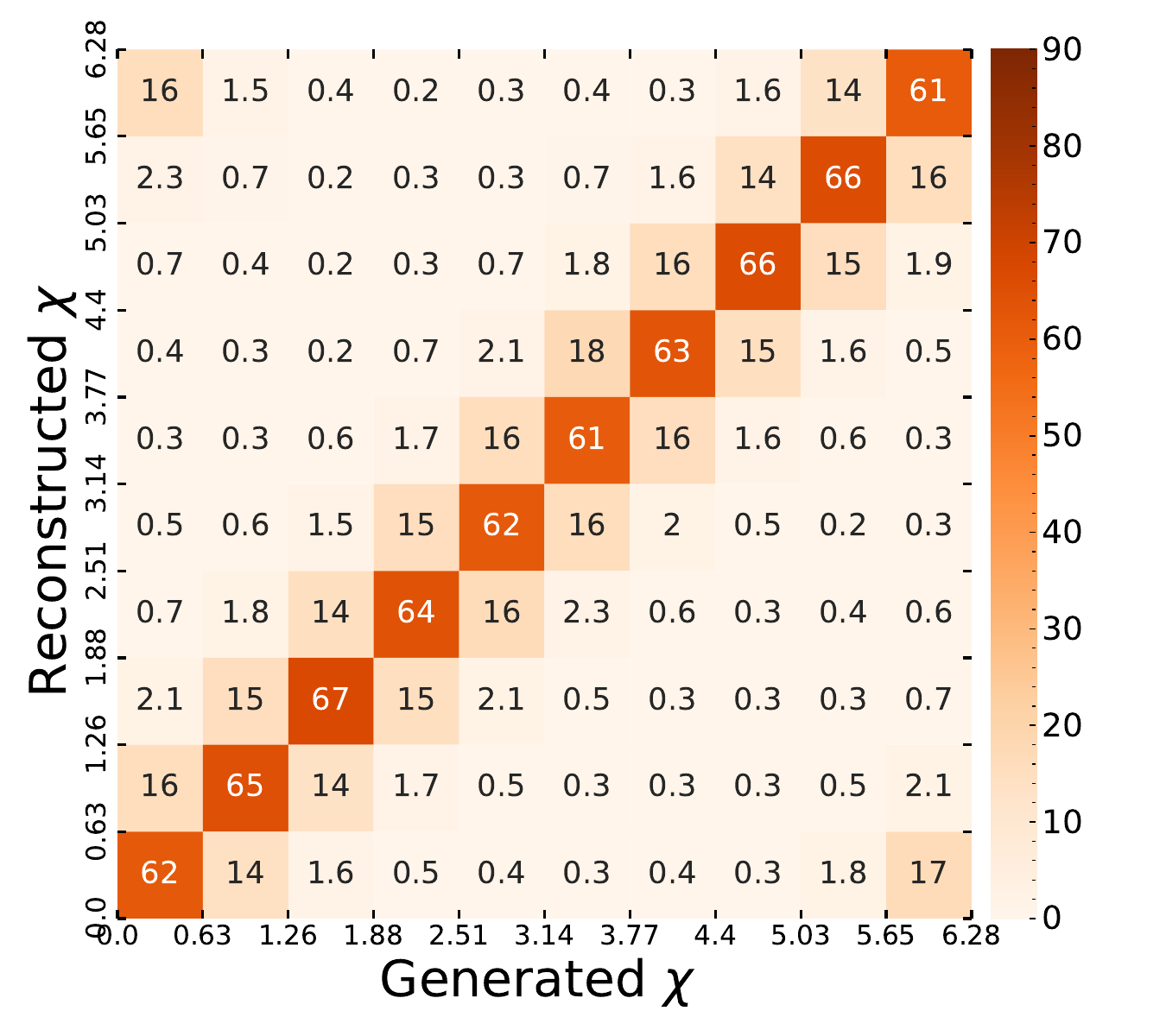}}
    \caption{Migration matrices of the reconstructed kinematic variables in the \bdsmunu decay. The values in the figures are in units of $10^{-2}$.} \label{fig:migration mu}
\end{figure*}

The migration matrices of \kinematic for the \bdsmunu decay are shown in Fig.~\ref{fig:migration mu}.

\clearpage
\FloatBarrier
\section{CORRELATIONS OF PARTIAL DECAY RATE }\label{app:correlation}

\begin{figure}[h!]
    \centering
    \subfigure{\includegraphics[width=0.72\textwidth, trim = 0.0cm 0cm 0cm 0.0cm, clip]{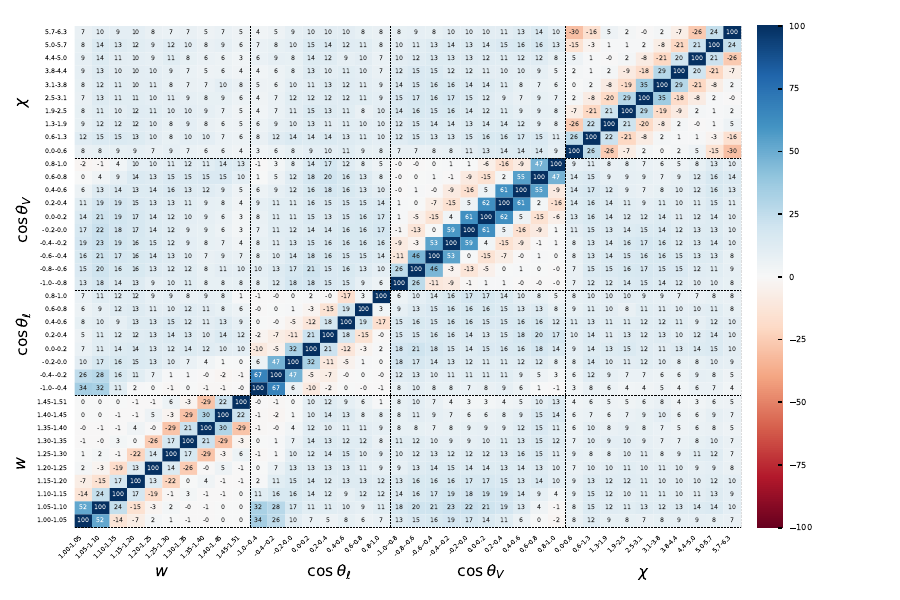}}\\
    \subfigure{\includegraphics[width=0.72\textwidth, trim = 0.0cm 0cm 0cm 0.0cm, clip]{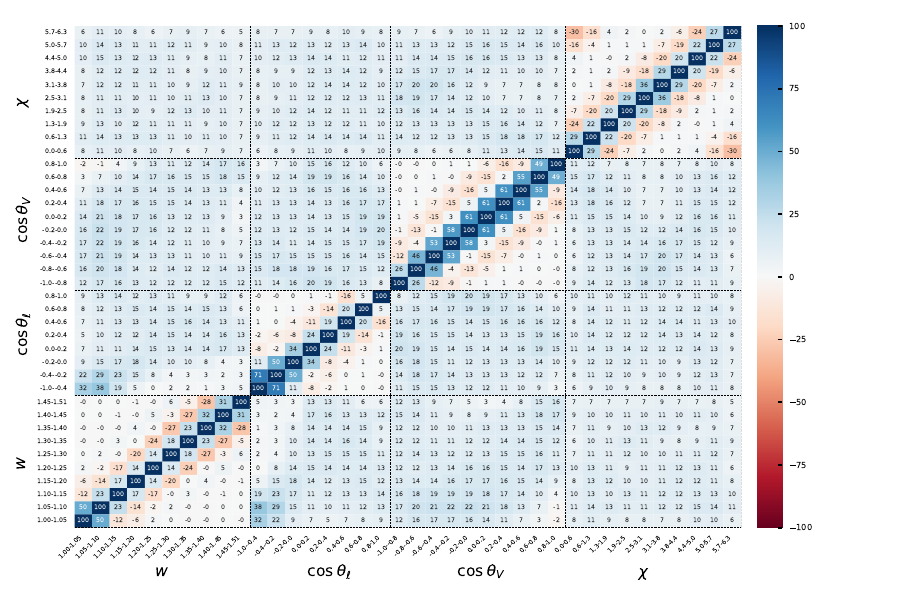}}
    \caption{Statistical correlations (in \%) of the partial decay rates for the \bdsenu (top) and \bdsmunu (bottom) decays.}
    \label{fig:decay rate statistical correlation}
\end{figure}

\begin{figure}
    \centering
    \subfigure{\includegraphics[width=0.72\textwidth, trim = 0.0cm 0cm 0cm 0.0cm, clip]{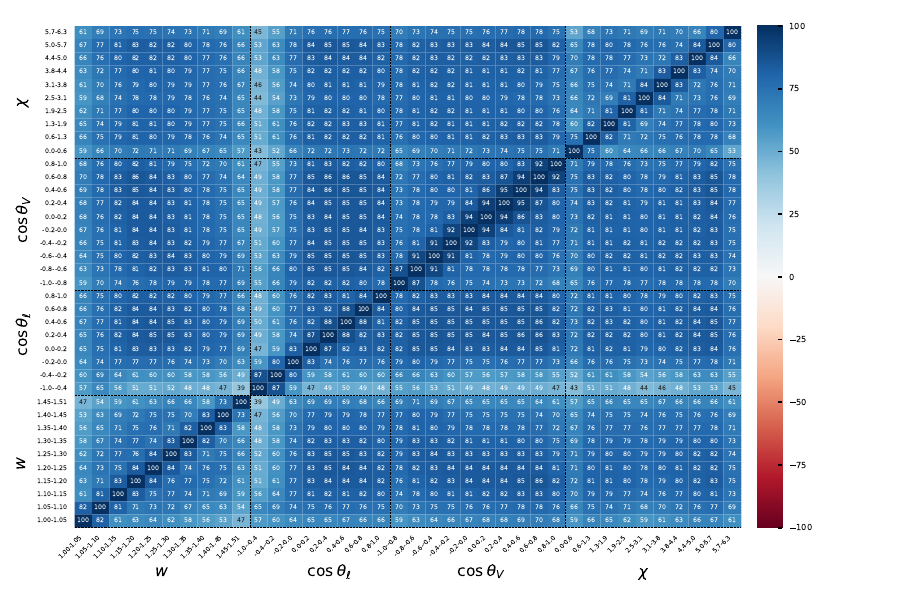}}\\
    \subfigure{\includegraphics[width=0.72\textwidth, trim = 0.0cm 0cm 0cm 0.0cm, clip]{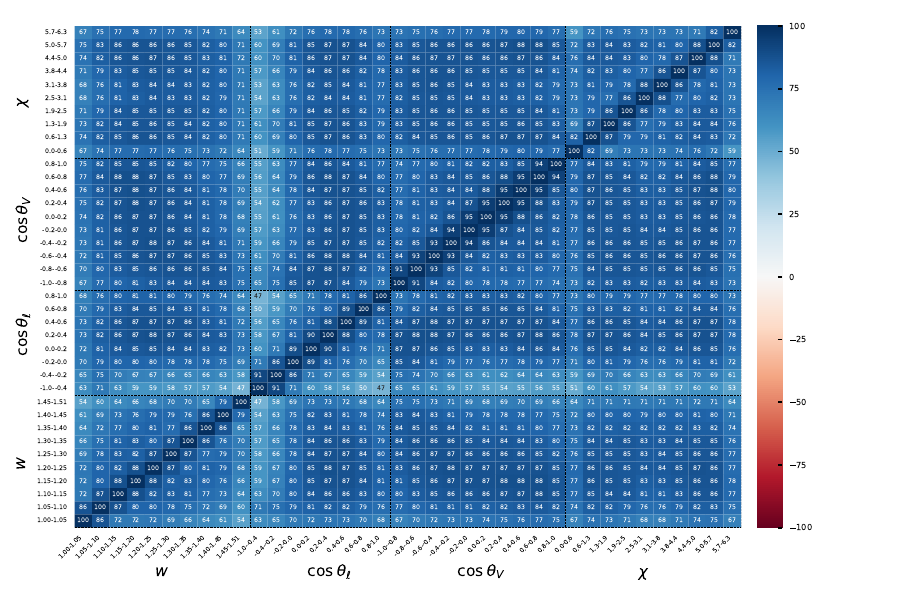}}
    \caption{Full experimental (statistical and systematic) correlations (in \%) of the partial decay rates for the \bdsenu (top) and \bdsmunu (bottom) decays.}
    \label{fig:decay rate correlation}
\end{figure}

\begin{figure}
    \centering
    \subfigure{\includegraphics[width=0.72\textwidth, trim = 0.0cm 0cm 0cm 0.0cm, clip]{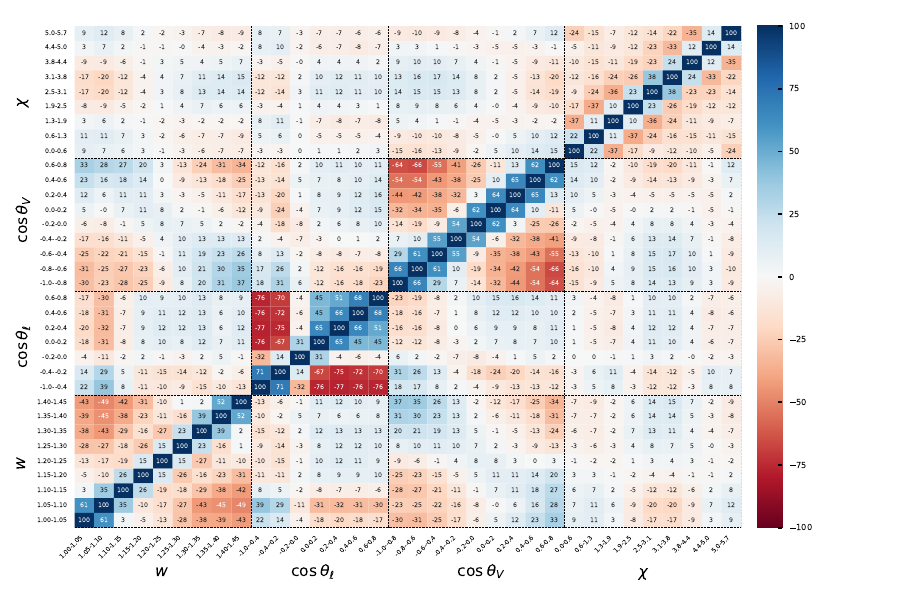}}\\
    \caption{Full experimental (statistical and systematic) correlations (in \%) for the average of the normalized partial decay rates. The last bin of each projection is excluded in the determination of \Vcb value, thus it is not shown.}
    \label{fig:normalized decay rate correlation}
\end{figure}

The statistical and full experimental correlations of the partial decay rates are provided in Figs.~\ref{fig:decay rate statistical correlation} and \ref{fig:decay rate correlation}, respectively. The full experimental correlations for the average of normalized partial decay rates over \bdsenu and \bdsmunu decays are given in Fig.~\ref{fig:normalized decay rate correlation}.

\FloatBarrier
\section{STATISTICAL AND SYSTEMATIC UNCERTAINTIES }\label{app:uncertainty}

The uncertainties on the partial decay rates in bins of the \kinematic projections are classified and summarized in Tables~\ref{tab:e uncertainty} and \ref{tab:mu uncertainty} for the \bdsenu and \bdsmunu decays, respectively.

\begin{sidewaystable}
% \tiny
\centering
\caption{Fractional uncertainties (in \%) of the partial decay rate in each bin for the \bdsenu decay.}\label{tab:e uncertainty}
\begin{tabular}{clccccccccccccc}
\hline\hline
                          Variable & \multicolumn{1}{c}{~Bin~} &  ~Statistical~ &  ~\makecell{Simulated \\sample size}~ &  ~\makecell{Signal\\ modeling}~ &  ~\makecell{Background\\ substraction}~ &  ~\makecell{Lepton ID\\ efficiency}~ &  ~\makecell{Slow-pion\\ efficiency}~ &  ~\makecell{Tracking of\\ $K$, $\pi$, $\ell$}~ &  ~$N_{B\Bbar}$~ &  ~$f_{+0}$~ &  ${\cal B}(D^{*}\rightarrow D\pi)$ &  ${\cal B}(D^{0}\rightarrow K\pi)$ &  ~\makecell{$B^0$\\ lifetime} \\
\hline
              \multirow{10}{*}{$w$} &              $[1.00, 1.05)$ &           3.56 &                              1.48 &                             0.57 &                                    2.53 &                                 0.65 &                                 4.24 &                                           0.90 &            1.52 &        2.52 &                               0.74 &                               0.76 &                          0.26 \\
                                    &              $[1.05, 1.10)$ &           2.25 &                              0.96 &                             0.26 &                                    1.72 &                                 0.53 &                                 3.72 &                                           0.90 &            1.52 &        2.52 &                               0.74 &                               0.76 &                          0.26 \\
                                    &              $[1.10, 1.15)$ &           1.94 &                              0.82 &                             0.55 &                                    1.27 &                                 0.51 &                                 3.34 &                                           0.90 &            1.52 &        2.52 &                               0.74 &                               0.76 &                          0.26 \\
                                    &              $[1.15, 1.20)$ &           1.74 &                              0.74 &                             0.87 &                                    1.07 &                                 0.45 &                                 3.17 &                                           0.90 &            1.52 &        2.52 &                               0.74 &                               0.76 &                          0.26 \\
                                    &              $[1.20, 1.25)$ &           1.72 &                              0.70 &                             0.80 &                                    0.98 &                                 0.44 &                                 2.88 &                                           0.90 &            1.52 &        2.52 &                               0.74 &                               0.76 &                          0.26 \\
                                    &              $[1.25, 1.30)$ &           1.75 &                              0.73 &                             0.70 &                                    0.87 &                                 0.46 &                                 2.73 &                                           0.90 &            1.52 &        2.52 &                               0.74 &                               0.76 &                          0.26 \\
                                    &              $[1.30, 1.35)$ &           1.81 &                              0.84 &                             0.94 &                                    0.85 &                                 0.45 &                                 2.54 &                                           0.90 &            1.52 &        2.52 &                               0.74 &                               0.76 &                          0.26 \\
                                    &              $[1.35, 1.40)$ &           1.94 &                              0.78 &                             1.05 &                                    0.83 &                                 0.46 &                                 2.27 &                                           0.90 &            1.52 &        2.52 &                               0.74 &                               0.76 &                          0.26 \\
                                    &              $[1.40, 1.45)$ &           2.02 &                              0.85 &                             0.92 &                                    0.80 &                                 0.47 &                                 2.10 &                                           0.90 &            1.52 &        2.52 &                               0.74 &                               0.76 &                          0.26 \\
                                    &              $[1.45, 1.51)$ &           2.97 &                              1.36 &                             1.27 &                                    0.24 &                                 0.51 &                                 1.93 &                                           0.90 &            1.52 &        2.52 &                               0.74 &                               0.76 &                          0.26 \\ \hline
 \multirow{8}{*}{$\cos\theta_\ell$} &            $[-1.00, -0.40)$ &           3.58 &                              1.40 &                             3.96 &                                    1.99 &                                 0.53 &                                 3.57 &                                           0.90 &            1.52 &        2.52 &                               0.74 &                               0.76 &                          0.26 \\
                                    &            $[-0.40, -0.20)$ &           2.50 &                              1.02 &                             3.08 &                                    1.44 &                                 0.62 &                                 3.27 &                                           0.90 &            1.52 &        2.52 &                               0.74 &                               0.76 &                          0.26 \\
                                    &             $[-0.20, 0.00)$ &           1.95 &                              0.80 &                             1.23 &                                    1.18 &                                 0.58 &                                 2.88 &                                           0.90 &            1.52 &        2.52 &                               0.74 &                               0.76 &                          0.26 \\
                                    &              $[0.00, 0.20)$ &           1.56 &                              0.62 &                             0.66 &                                    0.97 &                                 0.56 &                                 2.59 &                                           0.90 &            1.52 &        2.52 &                               0.74 &                               0.76 &                          0.26 \\
                                    &              $[0.20, 0.40)$ &           1.37 &                              0.61 &                             0.66 &                                    0.91 &                                 0.62 &                                 2.62 &                                           0.90 &            1.52 &        2.52 &                               0.74 &                               0.76 &                          0.26 \\
                                    &              $[0.40, 0.60)$ &           1.37 &                              0.54 &                             0.59 &                                    0.93 &                                 0.64 &                                 2.67 &                                           0.90 &            1.52 &        2.52 &                               0.74 &                               0.76 &                          0.26 \\
                                    &              $[0.60, 0.80)$ &           1.41 &                              0.58 &                             0.76 &                                    0.95 &                                 0.46 &                                 2.73 &                                           0.90 &            1.52 &        2.52 &                               0.74 &                               0.76 &                          0.26 \\
                                    &              $[0.80, 1.00)$ &           1.54 &                              0.68 &                             0.81 &                                    1.10 &                                 0.30 &                                 2.79 &                                           0.90 &            1.52 &        2.52 &                               0.74 &                               0.76 &                          0.26 \\ \hline
   \multirow{10}{*}{$\cos\theta_V$} &            $[-1.00, -0.80)$ &           1.53 &                              0.70 &                             0.47 &                                    0.89 &                                 0.53 &                                 2.07 &                                           0.90 &            1.52 &        2.52 &                               0.74 &                               0.76 &                          0.26 \\
                                    &            $[-0.80, -0.60)$ &           1.36 &                              0.58 &                             0.42 &                                    0.86 &                                 0.52 &                                 2.27 &                                           0.90 &            1.52 &        2.52 &                               0.74 &                               0.76 &                          0.26 \\
                                    &            $[-0.60, -0.40)$ &           1.49 &                              0.66 &                             0.64 &                                    1.01 &                                 0.50 &                                 2.54 &                                           0.90 &            1.52 &        2.52 &                               0.74 &                               0.76 &                          0.26 \\
                                    &            $[-0.40, -0.20)$ &           1.60 &                              0.68 &                             0.91 &                                    1.04 &                                 0.47 &                                 2.65 &                                           0.90 &            1.52 &        2.52 &                               0.74 &                               0.76 &                          0.26 \\
                                    &             $[-0.20, 0.00)$ &           1.66 &                              0.72 &                             1.24 &                                    0.98 &                                 0.45 &                                 2.85 &                                           0.90 &            1.52 &        2.52 &                               0.74 &                               0.76 &                          0.26 \\
                                    &              $[0.00, 0.20)$ &           1.67 &                              0.71 &                             1.44 &                                    0.87 &                                 0.45 &                                 3.07 &                                           0.90 &            1.52 &        2.52 &                               0.74 &                               0.76 &                          0.26 \\
                                    &              $[0.20, 0.40)$ &           1.66 &                              0.72 &                             1.46 &                                    0.77 &                                 0.44 &                                 3.17 &                                           0.90 &            1.52 &        2.52 &                               0.74 &                               0.76 &                          0.26 \\
                                    &              $[0.40, 0.60)$ &           1.61 &                              0.67 &                             1.38 &                                    0.66 &                                 0.46 &                                 3.40 &                                           0.90 &            1.52 &        2.52 &                               0.74 &                               0.76 &                          0.26 \\
                                    &              $[0.60, 0.80)$ &           1.51 &                              0.63 &                             1.24 &                                    0.54 &                                 0.45 &                                 3.55 &                                           0.90 &            1.52 &        2.52 &                               0.74 &                               0.76 &                          0.26 \\
                                    &              $[0.80, 1.00)$ &           1.98 &                              0.86 &                             1.18 &                                    0.12 &                                 0.49 &                                 3.65 &                                           0.90 &            1.52 &        2.52 &                               0.74 &                               0.76 &                          0.26 \\ \hline
           \multirow{10}{*}{$\chi$} &              $[0.00, 0.63)$ &           3.02 &                              1.31 &                             1.02 &                                    1.87 &                                 0.46 &                                 3.05 &                                           0.90 &            1.52 &        2.52 &                               0.74 &                               0.76 &                          0.26 \\
                                    &              $[0.63, 1.26)$ &           1.96 &                              0.82 &                             0.87 &                                    1.29 &                                 0.46 &                                 2.89 &                                           0.90 &            1.52 &        2.52 &                               0.74 &                               0.76 &                          0.26 \\
                                    &              $[1.26, 1.88)$ &           1.87 &                              0.81 &                             0.76 &                                    1.20 &                                 0.46 &                                 2.73 &                                           0.90 &            1.52 &        2.52 &                               0.74 &                               0.76 &                          0.26 \\
                                    &              $[1.88, 2.51)$ &           1.96 &                              0.83 &                             0.86 &                                    1.22 &                                 0.51 &                                 2.62 &                                           0.90 &            1.52 &        2.52 &                               0.74 &                               0.76 &                          0.26 \\
                                    &              $[2.51, 3.14)$ &           2.06 &                              0.91 &                             1.10 &                                    1.19 &                                 0.49 &                                 2.44 &                                           0.90 &            1.52 &        2.52 &                               0.74 &                               0.76 &                          0.26 \\
                                    &              $[3.14, 3.77)$ &           2.03 &                              0.85 &                             1.06 &                                    1.02 &                                 0.51 &                                 2.55 &                                           0.90 &            1.52 &        2.52 &                               0.74 &                               0.76 &                          0.26 \\
                                    &              $[3.77, 4.40)$ &           1.98 &                              0.87 &                             0.88 &                                    0.88 &                                 0.51 &                                 2.62 &                                           0.90 &            1.52 &        2.52 &                               0.74 &                               0.76 &                          0.26 \\
                                    &              $[4.40, 5.03)$ &           1.92 &                              0.78 &                             0.67 &                                    0.76 &                                 0.51 &                                 2.77 &                                           0.90 &            1.52 &        2.52 &                               0.74 &                               0.76 &                          0.26 \\
                                    &              $[5.03, 5.65)$ &           1.89 &                              0.80 &                             0.72 &                                    0.63 &                                 0.46 &                                 2.93 &                                           0.90 &            1.52 &        2.52 &                               0.74 &                               0.76 &                          0.26 \\
                                    &              $[5.65, 6.28)$ &           2.89 &                              1.20 &                             1.04 &                                    0.51 &                                 0.43 &                                 2.94 &                                           0.90 &            1.52 &        2.52 &                               0.74 &                               0.76 &                          0.26 \\
\hline\hline
\end{tabular}
\end{sidewaystable}

\begin{sidewaystable}
\centering
\caption{Fractional uncertainties (in \%) of the partial decay rate in each bin for the \bdsmunu decay.}\label{tab:mu uncertainty}
\begin{tabular}{clcccccccccccccc}
\hline\hline
                          Variable & \multicolumn{1}{c}{~Bin~} &  ~Statistical~ &  ~\makecell{Simulated \\sample size}~ &  ~\makecell{Signal\\ modeling}~ &  ~\makecell{Background\\ subtraction}~ &  ~\makecell{Lepton ID\\ efficiency}~ &  ~\makecell{Slow-pion\\ efficiency}~ &  ~\makecell{Tracking of\\ $K$, $\pi$, $\ell$}~ &  ~$N_{B\Bbar}$~ &  ~$f_{+0}$~ &  ${\cal B}(D^{*}\rightarrow D\pi)$ &  ${\cal B}(D^{0}\rightarrow K\pi)$ &  ~\makecell{$B^0$\\ lifetime} \\
\hline
              \multirow{10}{*}{$w$} &              $[1.00, 1.05)$ &           3.35 &                              1.32 &                             0.49 &                                   1.46 &                                 2.00 &                                 4.28 &                                           0.90 &            1.52 &        2.52 &                               0.74 &                               0.76 &                          0.26 \\
                                    &              $[1.05, 1.10)$ &           2.04 &                              0.79 &                             0.30 &                                   0.97 &                                 1.91 &                                 3.73 &                                           0.90 &            1.52 &        2.52 &                               0.74 &                               0.76 &                          0.26 \\
                                    &              $[1.10, 1.15)$ &           1.74 &                              0.70 &                             0.69 &                                   0.78 &                                 1.69 &                                 3.33 &                                           0.90 &            1.52 &        2.52 &                               0.74 &                               0.76 &                          0.26 \\
                                    &              $[1.15, 1.20)$ &           1.64 &                              0.72 &                             0.87 &                                   0.67 &                                 1.62 &                                 3.08 &                                           0.90 &            1.52 &        2.52 &                               0.74 &                               0.76 &                          0.26 \\
                                    &              $[1.20, 1.25)$ &           1.61 &                              0.67 &                             0.80 &                                   0.61 &                                 1.58 &                                 2.87 &                                           0.90 &            1.52 &        2.52 &                               0.74 &                               0.76 &                          0.26 \\
                                    &              $[1.25, 1.30)$ &           1.67 &                              0.70 &                             0.84 &                                   0.63 &                                 1.53 &                                 2.64 &                                           0.90 &            1.52 &        2.52 &                               0.74 &                               0.76 &                          0.26 \\
                                    &              $[1.30, 1.35)$ &           1.72 &                              0.68 &                             0.96 &                                   0.62 &                                 1.57 &                                 2.50 &                                           0.90 &            1.52 &        2.52 &                               0.74 &                               0.76 &                          0.26 \\
                                    &              $[1.35, 1.40)$ &           1.85 &                              0.70 &                             1.04 &                                   0.59 &                                 1.63 &                                 2.24 &                                           0.90 &            1.52 &        2.52 &                               0.74 &                               0.76 &                          0.26 \\
                                    &              $[1.40, 1.45)$ &           1.99 &                              0.78 &                             1.00 &                                   0.66 &                                 1.72 &                                 2.06 &                                           0.90 &            1.52 &        2.52 &                               0.74 &                               0.76 &                          0.26 \\
                                    &              $[1.45, 1.51)$ &           2.96 &                              1.15 &                             1.55 &                                   0.16 &                                 1.81 &                                 1.86 &                                           0.90 &            1.52 &        2.52 &                               0.74 &                               0.76 &                          0.26 \\ \hline
 \multirow{8}{*}{$\cos\theta_\ell$} &            $[-1.00, -0.40)$ &           3.36 &                              1.28 &                             4.20 &                                   1.85 &                                 3.56 &                                 3.76 &                                           0.90 &            1.52 &        2.52 &                               0.74 &                               0.76 &                          0.26 \\
                                    &            $[-0.40, -0.20)$ &           2.37 &                              0.95 &                             2.99 &                                   1.35 &                                 3.42 &                                 3.14 &                                           0.90 &            1.52 &        2.52 &                               0.74 &                               0.76 &                          0.26 \\
                                    &             $[-0.20, 0.00)$ &           1.88 &                              0.74 &                             1.11 &                                   1.11 &                                 3.42 &                                 2.89 &                                           0.90 &            1.52 &        2.52 &                               0.74 &                               0.76 &                          0.26 \\
                                    &              $[0.00, 0.20)$ &           1.45 &                              0.61 &                             0.59 &                                   0.90 &                                 2.95 &                                 2.59 &                                           0.90 &            1.52 &        2.52 &                               0.74 &                               0.76 &                          0.26 \\
                                    &              $[0.20, 0.40)$ &           1.34 &                              0.52 &                             0.64 &                                   0.89 &                                 2.02 &                                 2.56 &                                           0.90 &            1.52 &        2.52 &                               0.74 &                               0.76 &                          0.26 \\
                                    &              $[0.40, 0.60)$ &           1.27 &                              0.55 &                             0.53 &                                   0.88 &                                 1.14 &                                 2.65 &                                           0.90 &            1.52 &        2.52 &                               0.74 &                               0.76 &                          0.26 \\
                                    &              $[0.60, 0.80)$ &           1.29 &                              0.56 &                             0.68 &                                   0.87 &                                 0.42 &                                 2.67 &                                           0.90 &            1.52 &        2.52 &                               0.74 &                               0.76 &                          0.26 \\
                                    &              $[0.80, 1.00)$ &           1.47 &                              0.59 &                             0.70 &                                   1.04 &                                 0.15 &                                 2.75 &                                           0.90 &            1.52 &        2.52 &                               0.74 &                               0.76 &                          0.26 \\ \hline
   \multirow{10}{*}{$\cos\theta_V$} &            $[-1.00, -0.80)$ &           1.43 &                              0.64 &                             0.50 &                                   0.43 &                                 2.04 &                                 2.04 &                                           0.90 &            1.52 &        2.52 &                               0.74 &                               0.76 &                          0.26 \\
                                    &            $[-0.80, -0.60)$ &           1.35 &                              0.55 &                             0.45 &                                   0.58 &                                 1.77 &                                 2.21 &                                           0.90 &            1.52 &        2.52 &                               0.74 &                               0.76 &                          0.26 \\
                                    &            $[-0.60, -0.40)$ &           1.48 &                              0.62 &                             0.66 &                                   0.68 &                                 1.56 &                                 2.46 &                                           0.90 &            1.52 &        2.52 &                               0.74 &                               0.76 &                          0.26 \\
                                    &            $[-0.40, -0.20)$ &           1.55 &                              0.64 &                             0.95 &                                   0.68 &                                 1.46 &                                 2.69 &                                           0.90 &            1.52 &        2.52 &                               0.74 &                               0.76 &                          0.26 \\
                                    &             $[-0.20, 0.00)$ &           1.60 &                              0.68 &                             1.23 &                                   0.67 &                                 1.32 &                                 2.89 &                                           0.90 &            1.52 &        2.52 &                               0.74 &                               0.76 &                          0.26 \\
                                    &              $[0.00, 0.20)$ &           1.58 &                              0.66 &                             1.44 &                                   0.63 &                                 1.34 &                                 3.07 &                                           0.90 &            1.52 &        2.52 &                               0.74 &                               0.76 &                          0.26 \\
                                    &              $[0.20, 0.40)$ &           1.56 &                              0.66 &                             1.46 &                                   0.57 &                                 1.38 &                                 3.14 &                                           0.90 &            1.52 &        2.52 &                               0.74 &                               0.76 &                          0.26 \\
                                    &              $[0.40, 0.60)$ &           1.56 &                              0.63 &                             1.40 &                                   0.52 &                                 1.56 &                                 3.30 &                                           0.90 &            1.52 &        2.52 &                               0.74 &                               0.76 &                          0.26 \\
                                    &              $[0.60, 0.80)$ &           1.50 &                              0.61 &                             1.27 &                                   0.46 &                                 1.82 &                                 3.46 &                                           0.90 &            1.52 &        2.52 &                               0.74 &                               0.76 &                          0.26 \\
                                    &              $[0.80, 1.00)$ &           1.94 &                              0.81 &                             1.21 &                                   0.07 &                                 2.03 &                                 3.60 &                                           0.90 &            1.52 &        2.52 &                               0.74 &                               0.76 &                          0.26 \\ \hline
           \multirow{10}{*}{$\chi$} &              $[0.00, 0.63)$ &           2.88 &                              1.20 &                             1.15 &                                   1.02 &                                 1.57 &                                 2.90 &                                           0.90 &            1.52 &        2.52 &                               0.74 &                               0.76 &                          0.26 \\
                                    &              $[0.63, 1.26)$ &           1.83 &                              0.74 &                             0.82 &                                   0.77 &                                 1.74 &                                 2.91 &                                           0.90 &            1.52 &        2.52 &                               0.74 &                               0.76 &                          0.26 \\
                                    &              $[1.26, 1.88)$ &           1.79 &                              0.77 &                             0.68 &                                   0.80 &                                 1.78 &                                 2.77 &                                           0.90 &            1.52 &        2.52 &                               0.74 &                               0.76 &                          0.26 \\
                                    &              $[1.88, 2.51)$ &           1.84 &                              0.75 &                             0.89 &                                   0.77 &                                 1.63 &                                 2.60 &                                           0.90 &            1.52 &        2.52 &                               0.74 &                               0.76 &                          0.26 \\
                                    &              $[2.51, 3.14)$ &           1.98 &                              0.84 &                             1.11 &                                   0.79 &                                 1.60 &                                 2.50 &                                           0.90 &            1.52 &        2.52 &                               0.74 &                               0.76 &                          0.26 \\
                                    &              $[3.14, 3.77)$ &           1.95 &                              0.79 &                             1.10 &                                   0.74 &                                 1.56 &                                 2.46 &                                           0.90 &            1.52 &        2.52 &                               0.74 &                               0.76 &                          0.26 \\
                                    &              $[3.77, 4.40)$ &           1.86 &                              0.80 &                             0.89 &                                   0.65 &                                 1.69 &                                 2.58 &                                           0.90 &            1.52 &        2.52 &                               0.74 &                               0.76 &                          0.26 \\
                                    &              $[4.40, 5.03)$ &           1.75 &                              0.69 &                             0.74 &                                   0.52 &                                 1.68 &                                 2.73 &                                           0.90 &            1.52 &        2.52 &                               0.74 &                               0.76 &                          0.26 \\
                                    &              $[5.03, 5.65)$ &           1.78 &                              0.72 &                             0.81 &                                   0.54 &                                 1.70 &                                 2.83 &                                           0.90 &            1.52 &        2.52 &                               0.74 &                               0.76 &                          0.26 \\
                                    &              $[5.65, 6.28)$ &           3.00 &                              1.18 &                             0.93 &                                   0.32 &                                 1.59 &                                 2.98 &                                           0.90 &            1.52 &        2.52 &                               0.74 &                               0.76 &                          0.26 \\
\hline\hline
\end{tabular}
\end{sidewaystable}

\clearpage

\FloatBarrier
\section{INPUT PARAMETERS }\label{app:parameters}

In Table \ref{tab:input parameter}, we summarize the input parameters used in the BGL parametrization and the determination of partial decay rates.

\begin{table}[htbp]
\begin{center}
\caption{Input parameters of this analysis. All common inputs except the value of $f_{+0}$ are taken from Ref.~\cite{Workman:2022ynf}. The input parameters for the BGL parametrization are taken from Ref.~\cite{Grinstein:2017nlq}.
}\label{tab:input parameter}
\begin{tabular}{lccccc}
\hline\hline
       \multicolumn{2}{c}{Common input} \\
\hline
$m_{B^0}$               &  $5.27963$  \GeV \\
$m_{D^*}$               &  $2.01026$  \GeV \\
$\tau_{B^0}$            &  $(1.519\pm 0.004)\times 10^{-12}$ s  \\
${\cal B}(\DstDpi)$     &  $0.677\pm 0.005$   \\
${\cal B}(\DKpi)$       &  $0.03947\pm 0.00030$   \\ 
$\eta_{EW}$             &  $1.0066$ \\
$G_F$                   &  $1.1663787\times 10^{-5}$ GeV$^{-2}(\hbar c)^3$ \\
$f_{+0}$                &  $1.065\pm 0.052$~\cite{Belle:2022hka} \\
\hline 
\multicolumn{2}{c}{BGL input} \\ 
\hline
$n_I$                   & $2.6$ \\
$m_c/m_b$               & $0.33$  \\
$\chi^T(+ 0.33)$        & $5.28\times10^{-4}~ \mathrm{GeV}^{-2}$ \\
$\chi^T(- 0.33)$        & $3.07\times10^{-4}~ \mathrm{GeV}^{-2}$ \\
Vector $B_c^*$ masses   & $6.337$ \GeV  \\
                        & $6.899$ \GeV \\
                        & $7.012$ \GeV \\
                        & $7.280$ \GeV \\
Axial vector $B_c^*$ masses & $6.730$ \GeV  \\
                        & $6.736$ \GeV \\
                        & $7.135$ \GeV \\
                        & $7.142$ \GeV \\
\hline\hline
\end{tabular}
\end{center}
\end{table}

\FloatBarrier
\section{NESTED HYPOTHESIS TEST }\label{app:NHT}

A nested hypothesis test is carried out to determine the truncation of the BGL form-factor expansion order. It starts from $n_a=1$, $n_b=1$, $n_c=2$ (note that the value of $c_0$ is determined from $b_0$ parameter via Eq.~(\ref{eq:c0})) to allow at least one degree of freedom from each contributing form factor. We require all correlations between form-factor parameters to be smaller than $95\%$, and $\Delta\chi^2=\chi^2_N-\chi^2_{N+1}>1$, when one of the expansion of $g(z)$, $f(z)$, or ${\cal F}_1(z)$ is extended to a higher order.

\subsection{Test without LQCD input}

In this scenario, we only fit experimental data without LQCD predictions. The fitted \Vcb values, minima of the $\chi^2$, and numbers of degrees of freedom for various BGL expansion orders are summarized in Table~\ref{tab:NHT 1}. The fitted form-factor parameters with the optimal expansion order $n_a=1$, $n_b=2$, and $n_c=2$ are summarized in the main text in Table~\ref{tab:fit BGL}.

\begin{table}[htbp]
\begin{center}
\sisetup{uncertainty-mode=separate}
\renewcommand\arraystretch{1.3} 
\caption{Summary of the nested hypothesis test without LQCD input. The $\rho_{\text{max}}$ column records the largest off-diagonal correlation coefficients. The optimal expansion order is highlighted in bold.
}\label{tab:NHT 1}
\sisetup{uncertainty-mode=separate}
\begin{tabular}{cS[table-format=2.2(1)]S[table-format=1.2]S[table-format=2.1]S[table-format=2]c}
\hline\hline
    $(n_a, n_b, n_c)$ & {$|V_{cb}|\times 10^3$} & {$\rho_{\text{max}}$} &     {$\chi^2$} &          {Ndf} &    {$p$ value} \\
\hline
          $(1, 1, 2)$ &              40.2\pm1.1 &                  0.43 &             40 &             32 &           16\% \\
          $(2, 1, 2)$ &              40.1\pm1.1 &                  0.97 &           38.6 &             31 &           16\% \\
 \textbf{$(1, 2, 2)$} &     \textbf{$40.6\pm1.2$} &         \textbf{0.57} &  \textbf{38.9} &  \textbf{31} &  \textbf{16\%} \\
          $(1, 1, 3)$ &              40.1\pm1.1 &                  0.96 &           39.5 &             31 &           14\% \\
          $(2, 2, 2)$ &              40.3\pm1.3 &                  0.99 &           38.6 &             30 &           13\% \\
          $(1, 3, 2)$ &              40.0\pm1.3 &                  0.98 &             38 &             30 &           15\% \\
          $(1, 2, 3)$ &              40.5\pm1.2 &                  0.96 &           38.8 &             30 &           13\% \\
\hline\hline
\end{tabular}
\end{center}
\end{table}

\subsection{Test with FNAL/MILC lattice results of $h_{A_1}$}

In this scenario, we fit experimental data and the FNAL/MILC predictions on $h_{A_1}(w)$ at $w=[1.03, 1.10, 1.17]$ simultaneously. The obtained \Vcb values, minima of the $\chi^2$, and numbers of degrees of freedom corresponding to various truncations are summarized in Table~\ref{tab:NHT 2}. $n_a=1$, $n_b=1$, and $n_c=2$ is determined as the optimal expansion order. The fitted parameters and their correlations are summarized in Table~\ref{tab:NHT2 par}.

\begin{table}[htbp]
\begin{center}
\sisetup{uncertainty-mode=separate}
\renewcommand\arraystretch{1.3} 
\caption{Summary of the nested hypothesis test when FNAL/MILC predictions on $h_{A_1}(w)$ are taken into account.
}\label{tab:NHT 2}
\begin{tabular}{cS[table-format=2.2(1)]S[table-format=1.2]S[table-format=2.1]S[table-format=2]c}
\hline\hline
    $(n_a, n_b, n_c)$ & {$|V_{cb}|\times 10^3$} & {$\rho_{\text{max}}$} &     {$\chi^2$} &          {Ndf} &    {$p$ value} \\
\hline
 \textbf{$(1, 1, 2)$} &     \textbf{$40.0\pm1.2$} &         \textbf{0.62} &  \textbf{40.1} &  \textbf{34} &  \textbf{22\%} \\
          $(2, 1, 2)$ &              40.0\pm1.2 &                  0.97 &           38.6 &             33 &           23\% \\
          $(1, 2, 2)$ &              40.3\pm1.2 &                  0.59 &           39.2 &             33 &           21\% \\
          $(1, 1, 3)$ &              40.0\pm1.2 &                  0.96 &           39.5 &             33 &           20\% \\
\hline\hline
\end{tabular}
\end{center}
\end{table}

\begin{table}[htbp]
\begin{center}
\sisetup{uncertainty-mode=separate}
\renewcommand\arraystretch{1.3} 
\caption{Fitted parameters and their correlations using the optimal BGL expansion determined with FNAL/MILC constraints on $h_{A_1}(w)$.
}\label{tab:NHT2 par}
\begin{tabular}{cS[table-format=-2.2(1)]S[table-format=-1.2]S[table-format=-1.2]S[table-format=-1.2]S[table-format=-1.2]S[table-format=-1.2]S[table-format=-1.2]}
\hline\hline
{} &    {Value} & \multicolumn{4}{c}{Correlation} \\
\hline
$|V_{cb}|\times 10^3$ &  40.0\pm1.2 &            1.00 & -0.36 & -0.62 & -0.19 \\
$a_0\times 10^3$      &  21.5\pm1.3 &        -0.36 &     1.00 &  0.31 &  0.51 \\
$b_0\times 10^3$      &  13.2\pm0.2 &        -0.62 &  0.31 &     1.00 & -0.02 \\
$c_1\times 10^3$      &  -0.5\pm0.6 &        -0.19 &  0.51 & -0.02 &     1.00 \\
\hline\hline
\end{tabular}
\end{center}
\end{table}

\subsection{Test with FNAL/MILC lattice results of $h_{A_1}$, $R_{1}$, and $R_{2}$}

In the third scenario, we fit experimental data and the FNAL/MILC predictions on $h_{A_1}(w)$, $R_1(w)$, and $R_2(w)$ at $w=[1.03, 1.10, 1.17]$ simultaneously. The obtained \Vcb values, minima of the $\chi^2$, and numbers of degrees of freedom with various truncations are summarized in Table~\ref{tab:NHT 3}. $n_a=1$, $n_b=3$, and $n_c=2$ is determined as the optimal expansion order. The corresponding fitted parameters and their correlations are summarized in Table~\ref{tab:NHT3 par}.

\begin{table}[htbp]
\begin{center}
\sisetup{uncertainty-mode=separate}
\renewcommand\arraystretch{1.3} 
\caption{Summary of the nested hypothesis test when FNAL/MILC predictions on $h_{A_1}(w)$, $R_1(w)$, and $R_2(w)$ are taken into account.
}\label{tab:NHT 3}
\begin{tabular}{cS[table-format=2.2(1)]S[table-format=1.2]S[table-format=2.1]S[table-format=2]c}
\hline\hline
    $(n_a, n_b, n_c)$ & {$|V_{cb}|\times 10^3$} & {$\rho_{\text{max}}$} &     {$\chi^2$} &        {Ndf} &    {$p$ value} \\
\hline
          $(1, 1, 2)$ &              38.3\pm1.1 &                  0.57 &           75.3 &           40 &          0.1\% \\
          $(2, 1, 2)$ &              39.2\pm1.1 &                  0.59 &           52.4 &           39 &            7\% \\
          $(1, 2, 2)$ &              38.3\pm1.1 &                  0.61 &           75.2 &           39 &          0.1\% \\
          $(1, 1, 3)$ &              38.5\pm1.1 &                  0.92 &           73.6 &           39 &          0.1\% \\
          $(3, 1, 2)$ &              39.5\pm1.1 &                  0.85 &           48.7 &           38 &           11\% \\
          $(2, 2, 2)$ &              39.2\pm1.1 &                  0.59 &           52.3 &           38 &            6\% \\
          $(2, 1, 3)$ &              39.4\pm1.1 &                  0.92 &           50.2 &           38 &            9\% \\
          $(4, 1, 2)$ &              39.4\pm1.1 &                  0.98 &           48.4 &           37 &           10\% \\
          $(3, 2, 2)$ &              39.3\pm1.1 &                  0.87 &           47.5 &           37 &           12\% \\
 \textbf{$(3, 1, 3)$} &     \textbf{$39.7\pm1.1$} &         \textbf{0.92} &  \textbf{45.4} &  \textbf{37} &  \textbf{16\%} \\
          $(4, 2, 2)$ &              39.2\pm1.1 &                  0.98 &           46.8 &           36 &           11\% \\
          $(3, 3, 2)$ &              39.3\pm1.1 &                  0.87 &           46.4 &           36 &           11\% \\
          $(3, 2, 3)$ &              39.6\pm1.2 &                  0.91 &           45.2 &           36 &           14\% \\
          $(4, 3, 2)$ &              39.3\pm1.1 &                  0.98 &             46 &           35 &           10\% \\
          $(3, 4, 2)$ &              39.3\pm1.1 &                  0.86 &           46.2 &           35 &           10\% \\
          $(3, 3, 3)$ &              39.6\pm1.2 &                  0.93 &             45 &           35 &           12\% \\
          $(4, 1, 3)$ &              39.7\pm1.1 &                  0.98 &           44.1 &           36 &           17\% \\
          $(3, 1, 4)$ &              39.7\pm1.1 &                  0.91 &           45.3 &           36 &           14\% \\
          $(2, 2, 3)$ &              39.5\pm1.2 &                  0.91 &           50.1 &           37 &            7\% \\
          $(2, 1, 4)$ &              39.4\pm1.1 &                  0.91 &           50.1 &           37 &            7\% \\
          $(1, 2, 3)$ &              38.5\pm1.1 &                  0.91 &           73.6 &           38 &          0.1\% \\
          $(1, 1, 4)$ &              38.5\pm1.1 &                  0.91 &           73.4 &           38 &          0.1\% \\
\hline\hline
\end{tabular}
\end{center}
\end{table}

\begin{table}[htbp]
\begin{center}
\sisetup{uncertainty-mode=separate, table-align-uncertainty=true}
\renewcommand\arraystretch{1.3} 
\caption{Fitted parameters and their correlations using the optimal BGL expansion determined with FNAL/MILC constraints on $h_{A_1}(w)$, $R_1(w)$, and $R_2(w)$.
}\label{tab:NHT3 par}
\begin{tabular}{cS[table-format=-2.2(4)]S[table-format=-1.2]S[table-format=-1.2]S[table-format=-1.2]S[table-format=-1.2]S[table-format=-1.2]S[table-format=-1.2]S[table-format=-1.2]}
\hline\hline
{} &      {Value} & \multicolumn{7}{c}{Correlation} \\
\hline
$|V_{cb}|\times 10^3$ &    39.7\pm1.1 &            1.00 & -0.16 &  0.03 & -0.11 & -0.61 & -0.16 &  0.11 \\
$a_0\times 10^3$      &    28.1\pm1.0 &        -0.16 &     1.00 &  -0.10 & -0.19 &  0.17 &  0.12 & -0.03 \\
$a_1\times 10^3$      &  -44.2\pm65.8 &         0.03 &  -0.10 &     1.00 & -0.85 & -0.04 & -0.08 &  0.11 \\
$a_2$                 &    -5.1\pm2.4 &        -0.11 & -0.19 & -0.85 &     1.00 &  0.11 &  0.12 & -0.12 \\
$b_0\times 10^3$      &    13.3\pm0.2 &        -0.61 &  0.17 & -0.04 &  0.11 &     1.00 &   0.10 & -0.12 \\
$c_1\times 10^3$      &    -2.7\pm1.3 &        -0.16 &  0.12 & -0.08 &  0.12 &   0.10 &     1.00 & -0.92 \\
$c_2\times 10^3$      &   50.8\pm27.7 &         0.11 & -0.03 &  0.11 & -0.12 & -0.12 & -0.92 &     1.00 \\
\hline\hline
\end{tabular}
\end{center}
\end{table}

\begin{figure*}
    \centering
    \subfigure{\includegraphics[width=0.45\textwidth, trim = 0.42cm 0.10cm 1.0cm 0.45cm, clip]{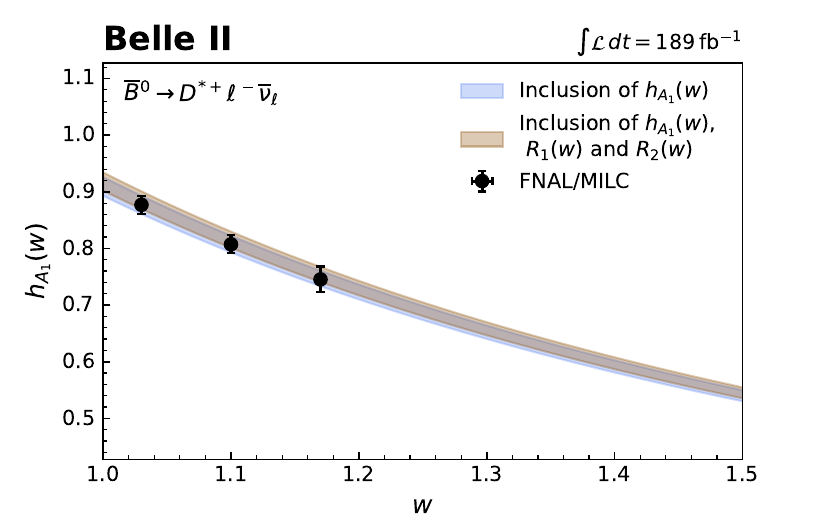}} \\
    \subfigure{\includegraphics[width=0.45\textwidth, trim = 0.42cm 0.10cm 1.0cm 0.45cm, clip]{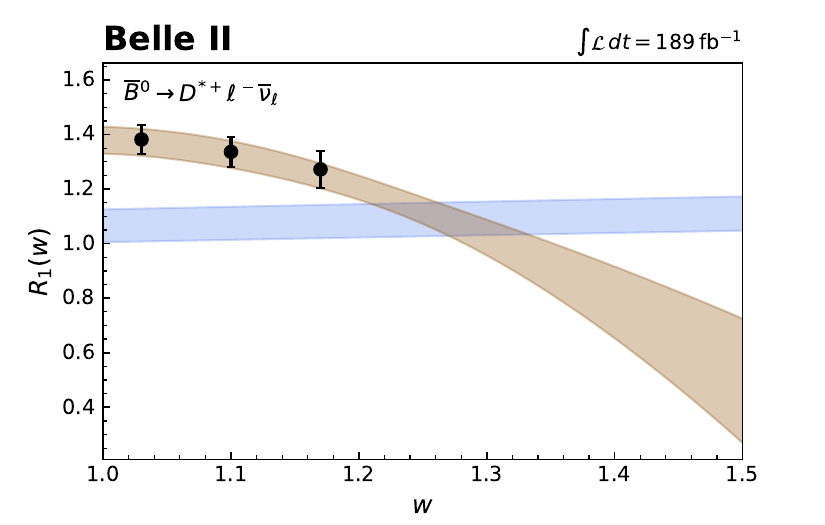}}
    \subfigure{\includegraphics[width=0.45\textwidth, trim = 0.42cm 0.10cm 1.0cm 0.45cm, clip]{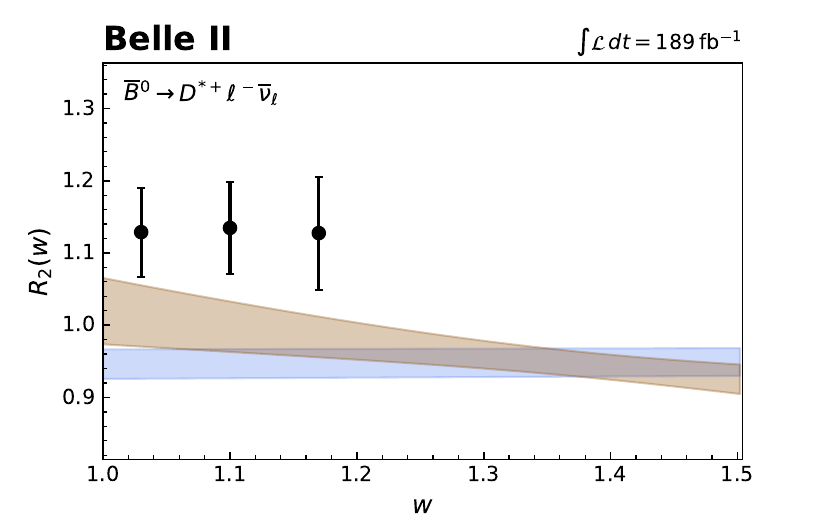}}
    \caption{Comparison of the $h_{A_1}(w)$, $R_1(w)$, and $R_2(w)$ spectra with the parameters determined in the nested hypothesis tests when FNAL/MILC lattice predictions are taken into account.}
    \label{fig:NHT}
\end{figure*}

By comparing results in the three scenarios, we find the inclusion of FNAL/MILC lattice results requires more BGL  form-factor parameters to reach an acceptable $\chi^2$ value. Using the fitted parameters summarized in Table~\ref{tab:NHT2 par} and Table~\ref{tab:NHT3 par}, we compare the $h_{A_1}(w)$, $R_1(w)$ and $R_2(w)$ spectra in Fig.~\ref{fig:NHT}. The reoptimized truncation results in a better description of lattice data, while the shapes of partial decay rates remain the same. 

\end{appendix}

%%%%%%%%%%%%%%%%%%%%%%%%%%%%%%%%%%%

\end{document}